\documentclass{article}
\usepackage{graphicx} 
\usepackage[a4paper, margin=1in]{geometry}
\usepackage{amsmath}
\usepackage{amssymb}
\usepackage{booktabs}
\usepackage[utf8]{inputenc}
\usepackage{cite}
\usepackage[super,sort&compress]{natbib}
\usepackage{url}
\usepackage{changepage}
\usepackage{tabularx}
\usepackage{caption}
\usepackage{placeins}
\usepackage{lineno}
\usepackage{centernot}

\title{Interpretable rainfall modelling reveals rapid reorganisation of Amazonian rainfall under vegetation loss}

\author{
  Lilly Horvath-Makkos\\
  \texttt{lilly.horvath-makkos@warwick.ac.uk}
  \and
  Fayyaz Minhas\\
  \texttt{fayyaz.minhas@warwick.ac.uk}
  \and
  Department of Computer Science, The University of Warwick
  \and Coventry, United Kingdom
}

\begin{document}
\date{}
\maketitle

\begin{abstract}
Understanding how vegetation loss alters rainfall remains a major challenge in climate and hydrological science, as deforestation modifies precipitation through heterogeneous, seasonal, and nonlinear land-atmosphere feedbacks. Existing modelling systems struggle to capture these dynamics: convection is parameterised at coarse spatial scales, potential tipping behaviour is poorly constrained, and rainfall–deforestation analyses are often limited to multi-decadal timescales. As a result, many approaches resolve correlations rather than causal effects, limiting our ability to anticipate hydrological disruption and inform water-security planning. Using a neural-network predictive model for hourly rainfall forecasting across the Amazon Basin, coupled with mechanistic pathway diagnostics and sensitivity analyses, we examine how vegetation perturbations reorganise rainfall dynamics across space, intensity regimes, and timescales. We test whether the model internalises physically consistent pathways linking vegetation, atmospheric state, and precipitation, and whether sustained canopy loss is associated with threshold behaviour in rainfall organisation. The model accurately predicts rainfall occurrence and intensity across the Amazon (Spearman = 0.84, F1 = 0.93, ROC-AUC = 0.98) and learns temporally ordered, physically consistent dependencies aligned with ecohydrological theory. From sensitivity analyses, we observe rapid and asymmetric rainfall responses to vegetation loss: heavy rainfall (20--50~mm~h$^{-1}$) declines by up to 7\% under sustained deforestation over eight months, while light rainfall (0.1--1~mm~h$^{-1}$) increases by nearly 4\%. Across scenarios, we observe rainfall entropy increases by 1.3\%, and dry-season intensity rises by 0.3--0.5\% per 0.5\% forest-cover loss, with the strongest disruptions occurring in the north-western Amazon and the Andean foothills. Through threshold analysis, we observe that after approximately 2--3 months of sustained vegetation changes in the most sensitive regions, the precipitating area fraction declines sharply. These findings demonstrate that data-driven methods can uncover process-relevant signatures of land–atmosphere coupling, offer new insights into hydrological vulnerability, and emphasise the urgency of Amazon conservation.
\end{abstract}

\newpage
\begin{figure}[t]
    \centering
    \includegraphics[width=\textwidth]{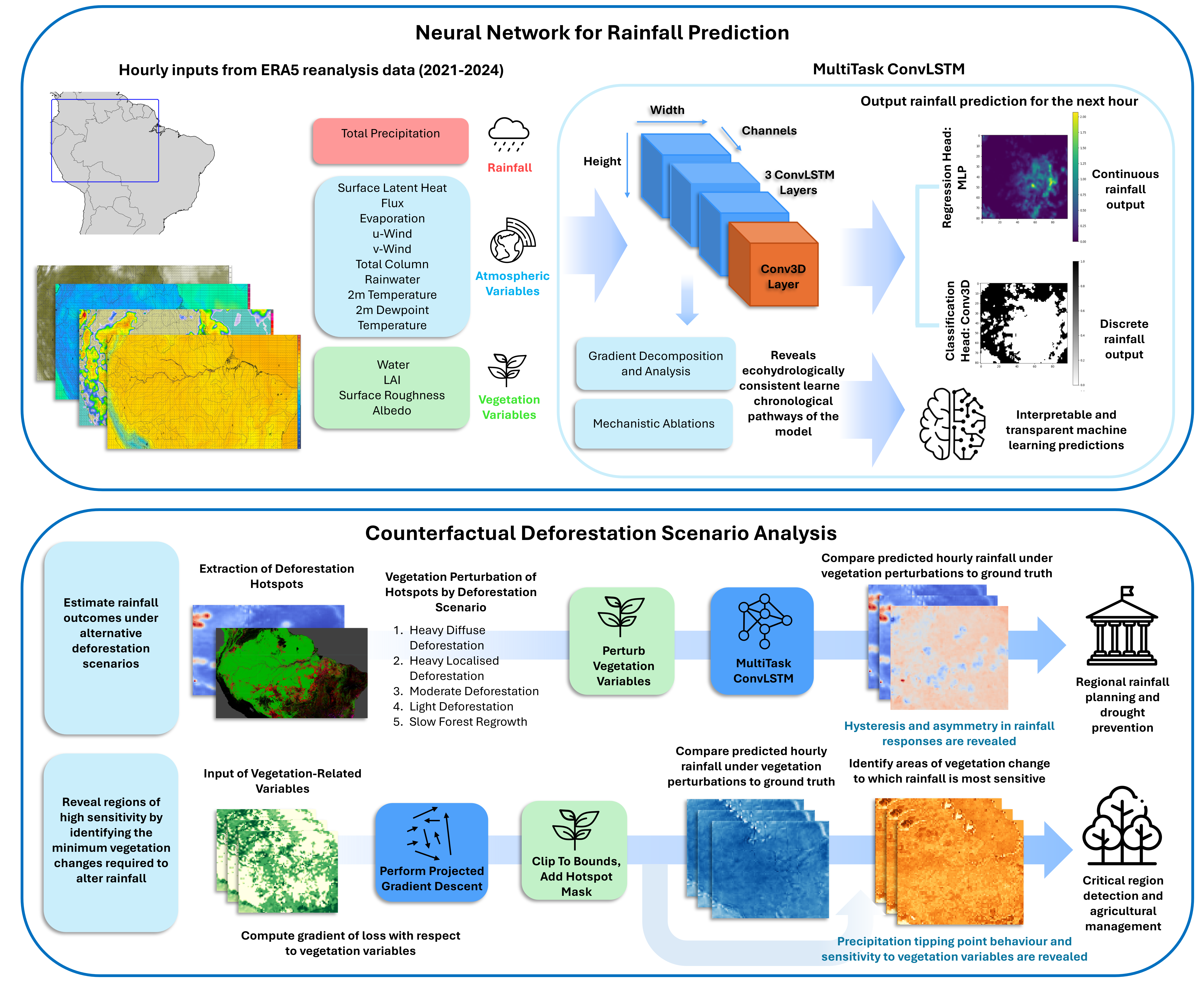}
    \caption*{\textbf{Graphical abstract:}  Interpretable framework for analysing vegetation-rainfall coupling over the Amazon Basin using ERA5 reanalysis data (2021–2024) at hourly resolution. A MultiTask ConvLSTM predicts rainfall occurrence and intensity from atmospheric and vegetation inputs (Spearman $\rho \approx 0.84$, F1 $\approx 0.93$, ROC-AUC $\approx 0.98$), enabling sensitivity-based attribution of land-atmosphere interactions consistent with established ecohydrological mechanisms. Counterfactual deforestation experiments reveal nonlinear and asymmetric rainfall responses, including reductions of up to 7\% in heavy rainfall (20–50~mm~h$^{-1}$), increases of up to 4\% in light rainfall (0–1~mm~h$^{-1}$), and an abrupt decline in precipitating area fraction after approximately 2–3 months of sustained vegetation loss along the most sensitive directions. Spatial sensitivity mapping identifies the north-western Amazon and Andean foothills as hotspots of short-term hydrological vulnerability.}
    \label{fig:graphical_abstract}
\end{figure}

\FloatBarrier

\newpage
\section{Introduction}

Tropical forests play a central role in regulating regional and global rainfall, with Amazonian deforestation disrupting hydrological cycles through reduced evapotranspiration, altered convection, and displacement of the intertropical convergence zone~\cite{bagley2014,badger2015,butt2023,konapala2020,lawrence2015}. These disruptions affect biodiversity, water security, and agriculture~\cite{mu2022,leitefilho2021}, and are increasingly linked to tipping points in both regional and global climate~\cite{duku2023,qin2025}. Currently, over 20\% of the Amazon rainforest is estimated to have been lost~\cite{dacruz2021}. As deforestation shows little sign of slowing to manageable levels, understanding short-term rainfall responses is increasingly urgent for mitigating local and regional climate impacts.

Rainfall regulation in tropical forests arises from tightly coupled exchanges of moisture, energy, and momentum between the land surface and atmosphere. Vegetation-related variables capture key pathways of land-atmosphere coupling: Leaf Area Index (LAI) controls evapotranspiration capacity and boundary-layer humidity~\cite{fang2019}; canopy and surface roughness regulate turbulent heat and momentum fluxes that influence boundary-layer depth and convective triggering~\cite{ma2019}; surface albedo alters radiative balance~\cite{zhang2022}; and soil water availability constrains latent heat flux, reinforcing drought through suppressed evapotranspiration~\cite{sanchezmartinez2025}. Operating across hours to months, these processes complicate attribution, as observed rainfall variability often reflects a combination of causal land-atmosphere feedbacks and coincident atmospheric variability. In the Amazon Basin, deforestation reduces latent heat flux, increases sensible heating, decreases roughness, and increases albedo, collectively suppressing convection and altering moisture convergence~\cite{lejeune2014}. Accurately capturing how these fast, localised processes generate rainfall extremes remains challenging, as many approaches smooth convective structure, misplace precipitation, underpredict high-intensity events, or lack mechanisms for attribution beyond correlation~\cite{Sonderby2020,30,li2025,xu2024,rui2024,liao2024short,wang2021multi,CNN_convective,4dvar}.

Rainfall responses to deforestation are also highly spatially heterogeneous, further limiting inference based on correlation alone. Southern Amazonia consistently exhibits drying in climate model experiments~\cite{bagley2014,medvigy2011}, reflecting its distance from Atlantic moisture sources and reliance on local recycling~\cite{qin2025}, whereas northwestern Amazonia often shows weaker sensitivity due to persistent Atlantic and Caribbean inflow~\cite{nian2024, dukuAssessingImpactsOngoing2023}. The Andes further modulate these responses by intensifying convection along the western margin while diverting low-level jets that redistribute moisture southward~\cite{chavez2017}. Large river systems such as the Amazon and Negro, together with coastal circulations, introduce additional sources of local moisture recycling, producing regions where rainfall may increase locally despite broader regional drying~\cite{lejeune2014}. As a result, spatial correlations between forest loss and rainfall can mask underlying feedbacks and yield opposing trends depending on scale. Sustained deforestation within concentrated hotspots may further induce hysteresis, whereby reduced evapotranspiration and enhanced sensible heating suppress convection beyond a threshold, limiting rainfall recovery even under regrowth~\cite{staal2020}. Identifying when and where such feedbacks emerge therefore, requires approaches that move beyond correlation toward causal reasoning at short timescales relevant for land-atmosphere coupling. This will assist in understanding the resilience of rainfall to disturbance.

Previous studies using global and regional climate models, including WRF-Noah~\cite{bagley2014} and OLAM~\cite{medvigy2011}, have documented substantial rainfall reductions in the southern Amazon under deforestation and revealed associated changes in moisture convergence and large-scale circulation. Observational analyses further suggest threshold behaviour, whereby prolonged forest loss in long-established hotspot regions ($>$10 years) leads to significant drying~\cite{mu2022}, while satellite-based estimates associate each 1\% forest loss with an average decline of 0.25~mm~month$^{-1}$ in rainfall~\cite{smith2023}. Together, these studies establish the sensitivity of Amazonian rainfall to vegetation loss, but leave open questions regarding short-term responses, spatially localised effects, and causal attribution. Moreover, numerical weather prediction and climate modelling frameworks typically rely on ensemble simulations and substantial computational resources, operate at coarse spatial or temporal resolution, and often struggle to accurately represent convective precipitation, limiting their ability to resolve fast, localised land-atmosphere feedbacks.

Recent advances in neural weather forecasting, including models such as ISOX~\cite{transformer} and 4DVarFormer~\cite{4dvar}, have demonstrated strong performance in large-scale precipitation estimation and extended-range forecasting. More broadly, machine learning (ML) approaches have gained prominence for precipitation nowcasting by learning spatiotemporal rainfall patterns directly from high-resolution reanalysis or radar data, complementing process-based climate and land-surface models. In particular, deep learning architectures such as Convolutional Long Short-Term Memory Networks (ConvLSTMs) and attention-based models have shown strong skill in predicting convective rainfall events~\cite{4dvar,21,sham2025,espeholt2022,Ravuri2021}, making them well suited for analysing short-term, localised precipitation dynamics and supporting efficient exploration of counterfactual scenarios.

Despite this potential, existing neural precipitation models remain limited in their ability to characterise how rainfall responds to underlying physical drivers. Recurrent architectures such as ConvLSTMs frequently underpredict extremes, struggle to distinguish between zero and non-zero rainfall, and exhibit skewed behaviour in high-intensity regimes~\cite{bagley2014,duku2023,CNN_convective,badger2015,butt2023,konapala2020,mu2022,leitefilho2021,qin2025,extreme_mean_precip,sham2025}. While the ConvLSTM architecture~\cite{21} and subsequent variants, including DConvLSTM-SAC~\cite{xu2024} and attention-based hybrids~\cite{wang2021multi,rui2024}, improve representation of spatial heterogeneity, these models are still predominantly evaluated in terms of predictive performance, with limited examination of whether their internal representations capture physically meaningful dependencies. Crucially, vegetation indices are rarely incorporated as dynamic drivers of precipitation, leaving land-atmosphere coupling weakly constrained and limiting mechanistic interpretation of rainfall responses to deforestation~\cite{30}.

Overall, no single existing modelling approach simultaneously resolves convective-scale dynamics at regional scales, treats vegetation as a dynamic forcing, supports counterfactual deforestation analysis, and identifies the dominant spatial-temporal drivers of rainfall variability while maintaining physically and chronologically consistent internal representations aligned with hydrological and ecohydrological theory. Consequently, key aspects of short-term vegetation-rainfall coupling in tropical forest systems remain insufficiently constrained.

In this study, we address these challenges by developing a modelling pipeline for high-resolution analysis of short-term precipitation responses to incremental and spatially concentrated vegetation perturbations using data-driven models. We quantify the magnitude, direction, and spatial variability of rainfall sensitivity to vegetation change, revealing non-linear responses, asymmetric behaviour under loss and regrowth, and the emergence of hysteresis. We further demonstrate that neural networks trained on observational data can internalise temporally consistent and directionally coherent dependencies between vegetation, atmospheric state, and rainfall. By incorporating mechanisms that enhance interpretability and transparency, we assess whether learned internal representations align with established ecohydrological pathways. This facilitates identification of variable-specific and location-dependent directions of rainfall sensitivity, providing new insight into the structure of land-atmosphere feedbacks at convective timescales and demonstrating how data-driven models can support interpretation of vegetation-driven precipitation change in tropical forest systems.

\vspace{0.5cm}
We summarise our main contributions as follows:

\begin{itemize}
    \item[(1)] We develop an interpretable neural forecasting model that jointly predicts hourly rainfall occurrence and intensity at convective timescales (Spearman $\rho \approx 0.84$, F1 $\approx 0.93$, ROC-AUC $\approx 0.98$), and show through targeted ablations and sensitivity analyses, that the model learns temporally ordered and physically consistent dependencies between vegetation changes and rainfall.
    
    \item[(2)] Using counterfactual deforestation experiments, we observe nonlinear rainfall redistributions under sustained canopy loss, with heavy rainfall (20--50~mm~h$^{-1}$) declining by up to 7\% and light rainfall (0--1~mm~h$^{-1}$) increasing by up to 4\% under heavy deforestation (forest loss $-1.0$ -- $-1.5\% ~\text{yr}^{-1}$).
    
    \item[(3)] We identify threshold-like behaviour in short-term rainfall organisation, whereby after approximately 2--3 months of worst-case sustained vegetation perturbation in the most sensitive regions, we observe an abrupt reduction in the precipitating area fraction.
    
    \item[(4)] We find that rainfall responses are strongly asymmetric, with deforestation suppressing extreme rainfall, while promoting light rainfall, while reforestation disproportionately increases extreme rainfall and rainfall entropy, indicating short-timescale hysteresis in vegetation-rainfall coupling.
    
    \item[(5)] We observe that rainfall sensitivity is highly spatially heterogeneous, with the north-western Amazon and Andean foothills exhibiting the strongest and most rapid disruptions, identifying locations of short-term hydrological vulnerability.
\end{itemize}

\section{Results}

We have developed a data-driven model capable of predicting hourly precipitation across the central Amazon using atmospheric and vegetation information from ERA5-Reanalysis (2021--2024). The model jointly predicts rainfall occurrence and intensity, representing both the timing and magnitude of precipitation events and capturing convective rainfall dynamics alongside land-atmosphere interactions at regional scales. We evaluate the model’s forecast fidelity and generalisation performance on two independent held-out test periods, including intervals coinciding with historical El Niño conditions that were not present during training, to establish whether the model can reliably reproduce observed short-term rainfall patterns under varying climatic regimes.

We next assess whether the model learns physically meaningful structure rather than relying on purely correlative relationships. To do so, we examine how rainfall predictions respond to targeted removal and perturbation of atmospheric and vegetation inputs, testing whether the model distinguishes between temporally ordered drivers of precipitation.

Finally, the trained model is applied to counterfactual scenarios representing deforestation and regrowth patterns. Resulting changes in rainfall characteristics are analysed to identify asymmetries between vegetation loss and recovery, spatially heterogeneous sensitivities, and the emergence of threshold behaviour in short-term rainfall response.

\subsection{Prediction of Hourly Rainfall Patterns}

\begin{table}
\centering
\small
\setlength{\tabcolsep}{5pt}
\renewcommand{\arraystretch}{1.05}
\caption{Precipitation prediction performance across models. The persistence baseline uses the previous hour’s precipitation as the prediction. The baseline ConvLSTM consists of three ConvLSTM layers. ConvLSTM + Conv3D extends this with an additional Conv3D output layer. The MultiTask ConvLSTM includes parallel regression and classification heads. Values are reported as mean over the test set. The dataset includes 43.65\% 0-precipitation values and 56.35\% non-zero precipitation.}
\label{tab:performance_metrics}
\begin{tabular}{lccccc}
\toprule
\multicolumn{6}{c}{\textbf{Regression and Correlation Metrics}} \\
\midrule
\textbf{Model} & \textbf{Spearman} & \textbf{Pearson} & \textbf{Kendall} & \textbf{NSE} & \textbf{MSE} \\
\midrule
Persistence Baseline     & $0.8012$ & $0.6728$ & $0.6806$ & $0.2641$ & $0.0350$\\
Baseline ConvLSTM        & $0.8230 $ & $0.8108 $ & $0.6580 $ & $0.6476 $ & $0.04151$\\
ConvLSTM + Conv3D        & $0.8388$ & $0.8231$ & $0.6690$ & $0.7035$ & $0.0413$\\
\textbf{MultiTask ConvLSTM} & \textbf{$0.8412$} & \textbf{$0.8302$} & \textbf{$0.6870 $} & \textbf{$0.6890 $} & \textbf{$0.0448$} \\
\midrule
\multicolumn{6}{c}{\textbf{Classification Metrics}} \\
\midrule
\textbf{Model} & \textbf{Accuracy} & \textbf{Precision} & \textbf{Recall} & \textbf{F1 Score} & \textbf{ROC-AUC} \\
\midrule
Persistence Baseline     & $0.8955 $ & $0.8913 $ & $0.8909$ & $0.8878 $  & $0.9012 $ \\
\textbf{MultiTask ConvLSTM} & \textbf{$0.9159 $} & \textbf{$0.9186 $} & \textbf{$0.9335 $} & \textbf{$0.9260 $} & \textbf{$0.9754 $} \\
\bottomrule
\end{tabular}
\end{table}

We evaluate the performance of the proposed multi-task convolutional long short-term memory network (MultiTask ConvLSTM) framework for predicting hourly precipitation across the Amazon basin. Model skill is assessed on independent test periods using grid-wise and basin-aggregated metrics at hourly resolution (see Methods). The MultiTask ConvLSTM learns spatial rainfall structures through convolutional filters while simultaneously tracking their temporal evolution via recurrent dynamics, enabling it to represent evolving storm systems and land-atmosphere feedbacks. The model achieves strong correlation-based metrics (Kendall Tau = 0.69, Spearman = 0.84), as shown in Table~\ref{tab:performance_metrics}, indicating accurate recovery of the structure of rainfall rank in space and time. These results suggest that subtle variations in precipitation intensity are preserved.

\begin{figure}[htbp!]
    \centering
    \includegraphics[width=1\linewidth]{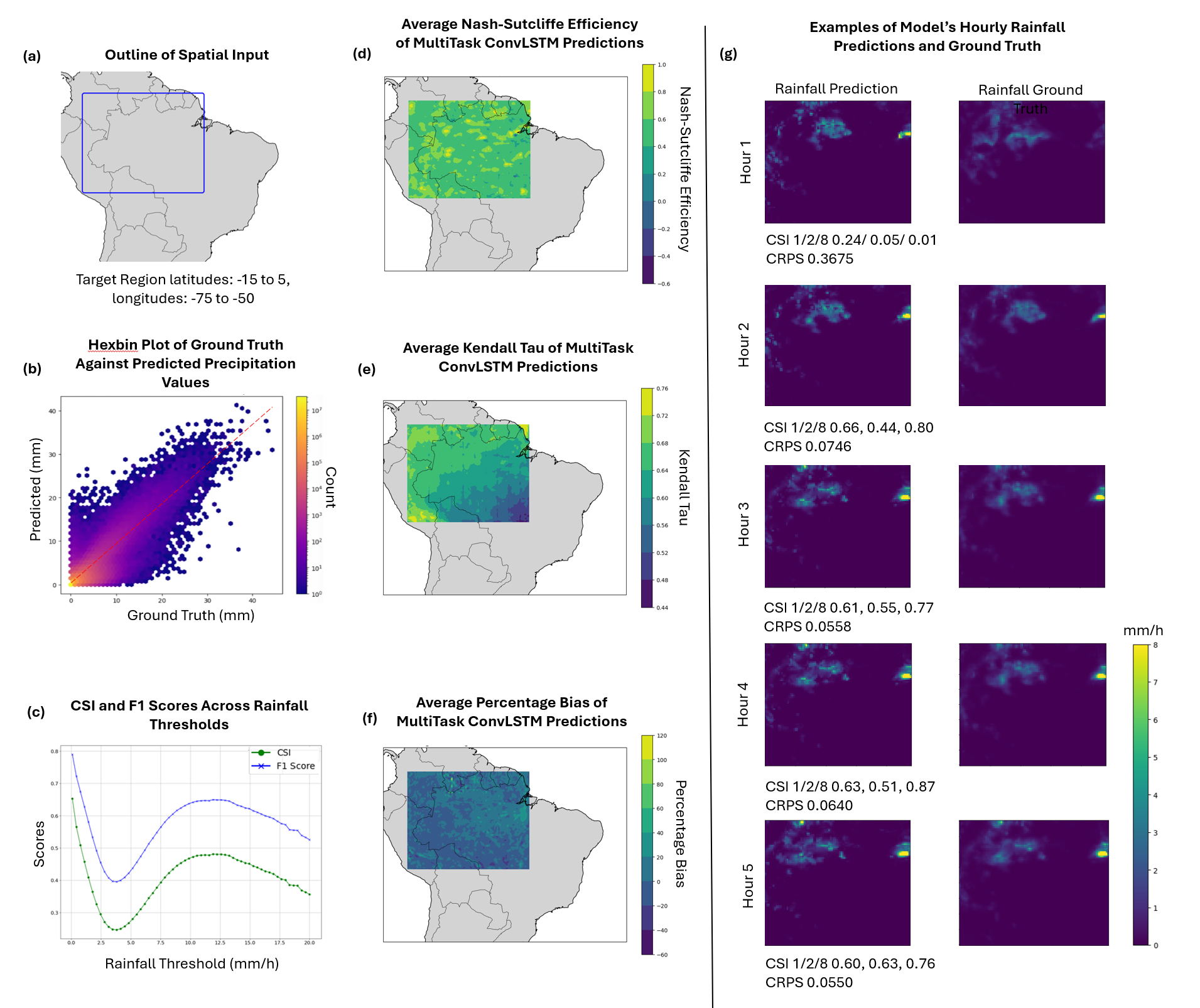}
    \caption{Visualisation of distribution and metrics of MultiTask ConvLSTM predictions across the test region. \textbf{a,} Spatial outline of the region of interest. Maps span latitude $-15^\circ$ to $5^\circ$ and longitude $-75^\circ$ to $-50^\circ$, and were generated in Python. \textbf{b,} Hexbin plot comparing predicted and ground truth precipitation values for the MultiTask ConvLSTM model. The red dashed line denotes the ideal 1:1 correspondence ($y = x$). Predictions closely align with the diagonal across the rainfall spectrum, indicating accurate model behaviour. \textbf{c,} Critical Success Index (CSI) and F1 score of the model compared to ERA5 ground truth across varying rainfall thresholds.\textbf{d,} Temporal average of hourly Nash-Sutcliffe Efficiency (NSE) for MultiTask ConvLSTM predictions on the test set. 
    \textbf{e,} Temporal average of hourly Kendall’s Tau correlation, indicating the model’s ability to preserve temporal rank ordering of rainfall intensities. 
    \textbf{f,} Temporal average hourly of Percentage Bias (PBIAS), measuring the systematic bias of the predictions.
    \textbf{g,} Hourly sequences of predicted rainfall (top), ground truth from the ERA5 Reanalysis dataset (bottom). Predictions are evaluated using the Critical Success Index (CSI) at thresholds of 1~mm/h, 2~mm/h, and 8~mm/h, alongside the Continuous Ranked Probability Score (CRPS). The model accurately captures the intensity, spatial structure, and temporal evolution of rainfall events, particularly for moderate rainfall.}
    \label{fig:model_performance}
\end{figure}

By jointly predicting a binary rainfall classification and continuous intensity estimate, the model handles dry-to-wet transitions effectively. Binary classification performance is strong (F1 = 0.93, ROC-AUC = 0.98), as shown in Table~\ref{tab:performance_metrics}, with over 90\% of rainfall onset and cessation events correctly identified across the region. Most misclassifications occur at low rainfall intensities ($<$1 mm/h), where atmospheric noise and labelling uncertainty are highest.

The MultiTask ConvLSTM achieves high fidelity in predicting rainfall distributions, closely aligning with observed patterns across intensities, as shown in Figure~\ref{fig:model_performance} \textbf{(b)}. This figure shows a dense concentration of predictions along the 1:1 line, with minimal skew. While a small number of samples over-predict low rainfall, the majority of predictions cluster accurately below 2mm/h, demonstrating strong dry-event precision. High-intensity rainfall events above 40~mm/h are also preserved, highlighting the model’s ability to capture extremes.

This performance is consistent with the CSI and CRPS scores in Figure~\ref{fig:model_performance} \textbf{(g)}. CSI values reach as high as 0.66 at the 1mm/h threshold and 0.80 at 8mm/h, indicating robust detection of moderate and extreme rainfall. CRPS values typically remain low, supporting accurate probabilistic forecasts and reinforcing the model’s precision in distinguishing wet-dry transitions and rainfall structure. There is only minor underprediction near convective edges and slight overestimation between 5-15~mm/h.

As further shown in Figure~\ref{fig:model_performance} \textbf{(d), (e), (f)}, performance is stable across space. Nash-Sutcliffe Efficiency (NSE) is generally high, though patchy in regions of convective complexity. Kendall’s Tau is strongest in the northwest and along the Andes, declining southward toward the Cerrado. This is because the southeastern Amazon and Cerrado transition zone has a climate influenced by convective initiation thresholds and synoptic variability, reducing the consistency of rainfall ranks across time and thus lowering Kendall's Tau. PBIAS remains close to zero, with no significant systematic over- or underestimation, except in a few urban regions. 

The CSI and F1 score curves  (Figure~\ref{fig:model_performance},\textbf{(f)}) exhibit a pronounced minimum at intermediate rainfall thresholds (2--5~mm/h), coinciding with the broadened density in the hexbin distribution, before recovering at higher intensities. This range corresponds to a transition regime between stratiform precipitation and organised deep convection, where rainfall is spatially fragmented, short-lived, and highly sensitive to small-scale triggering processes. At hourly resolution, such events are intrinsically difficult to classify, and modest temporal or spatial displacements are strongly penalised by categorical metrics. In contrast, light rainfall is more temporally persistent and spatially coherent, while heavy rainfall is more distinct once established, leading to higher skill at both ends of the intensity spectrum.

\FloatBarrier

\subsubsection{Model Learns Mechanistic Pathway from Vegetation to Rainfall}

\begin{table}
\centering
\small
\setlength{\tabcolsep}{4pt}
\renewcommand{\arraystretch}{1.05}
\caption{Ablation study results for the MultiTask ConvLSTM with vegetation (V), precipitation (P), and atmospheric (A) mediators. The full model includes all three input types (A, V, P). Comparing ablation patterns shows that removing precipitation memory strongly degrades performance, and that removing vegetation in addition to precipitation leads to a further decline, consistent with vegetation contributing predictive information via atmospheric mediation. Values are the mean across the test set. All results are reported to four decimal places.}
\label{tab:ablation_results}
\begin{tabularx}{\linewidth}{l *{5}{>{\centering\arraybackslash}X}}
\toprule
\textbf{Inputs} & \textbf{Spearman} & \textbf{Pearson} & \textbf{Kendall} & \textbf{NSE} & \textbf{MSE} \\
\midrule
A, P (no V)        & $0.8445$ & $0.8310$ & $0.6799$ & $0.6281 $ & $0.0223 $ \\
A only (no V, no P)& $0.6413 $ & $0.5404 $ & $0.4925 $ & $0.2827$ & $0.0210$ \\
A, V (no P)        & $0.6622 $ & $0.5669$ & $0.5092 $ & $0.3191 $ & $0.0199 $ \\
A, V, P (full)     & \textbf{$0.8412 $} & \textbf{$0.8302 $} & \textbf{$0.6870 $} & \textbf{$0.6890 $} & \textbf{$0.0448 $} \\
\bottomrule
\end{tabularx}
\end{table}

\begin{figure}[htbp!]
    \centering
    \includegraphics[width=\linewidth]{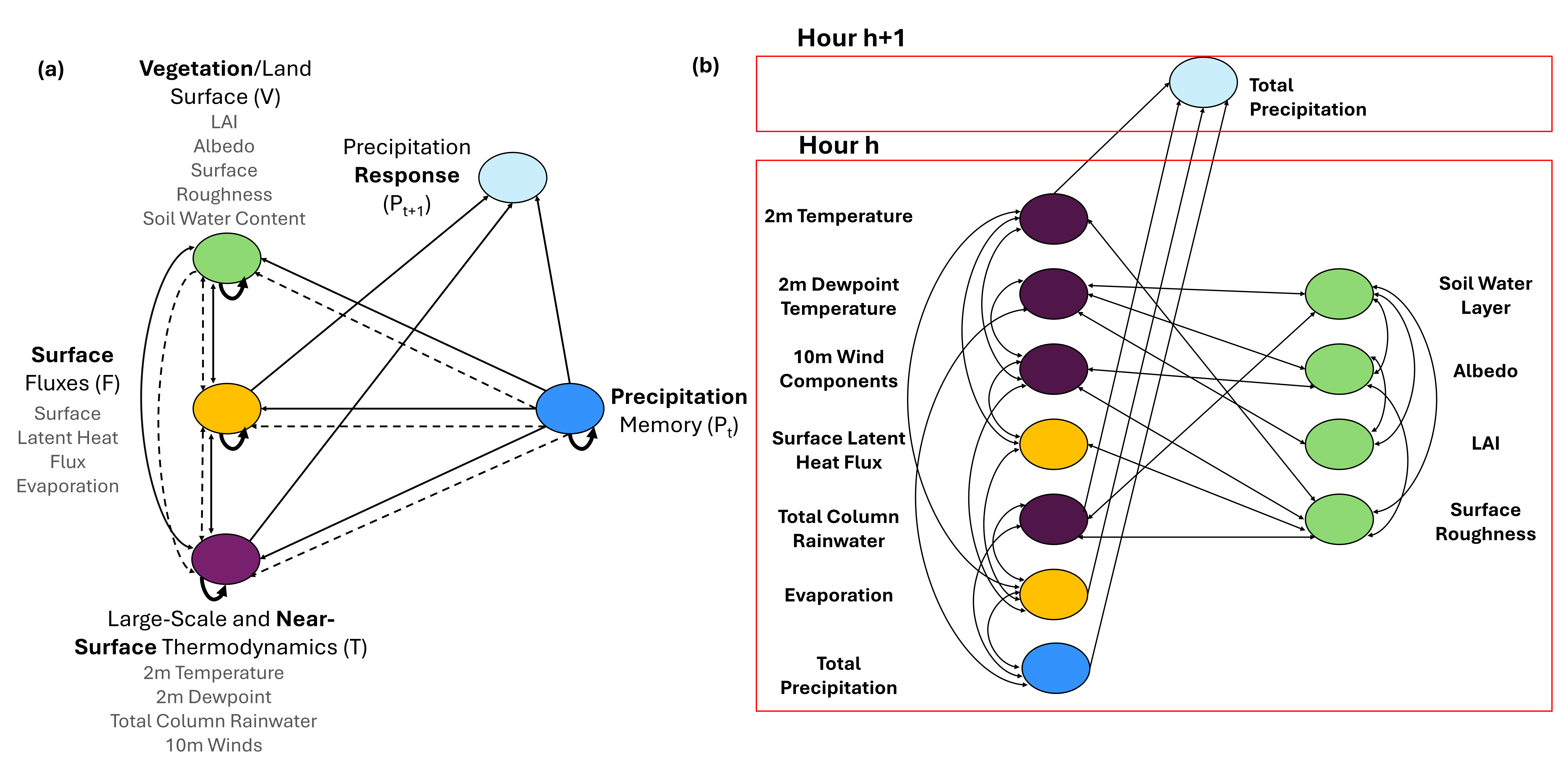}
    \caption{Two causal graphs showing the interactions between input variables. Green represents vegetation and land surface variables, including LAI, albedo, surface roughness, and soil water content. Orange represents surface fluxes, including surface latent heat flux and evaporation. Purple represents thermodynamic variables, including 2m temperature, 2m dewpoint temperature, total column rainwater, and wind components. Dark blue represents precipitation memory at the previous hour. Light blue represents precipitation at the next hour. \textbf{a, } Conceptual schematic of hypothesised land-atmosphere-precipitation pathways. Vegetation (V: LAI, albedo, roughness, soil water) influences fluxes (F: latent heat flux, evaporation) through stomatal control and surface structure, while fluxes feedback onto vegetation via soil moisture and energy balance. Fluxes and thermodynamics (T: temperature, dewpoint, rainwater, winds) interact through moistening, demand, and regime shifts. Vegetation and thermodynamics are linked via albedo-driven energy balance and climate stress. Precipitation memory (Pt-1) feeds back into all three groups: altering surface fluxes, thermodynamics, and vegetation (canopy stress and growth). Current precipitation (Pt) is directly driven by flux supply and thermodynamic instability, while autocorrelations (self-arrows) represent persistence within each group. Solid arrows denote fast, direct processes; dashed arrows indicate slower or indirect effects. \textbf{b, } Data-driven causal graph inferred using the PC algorithm, applied to model inputs at time $t-1$ and total precipitation at time $t$. Edges indicate statistically supported conditional dependencies under standard causal sufficiency assumptions. The inferred structure highlights an indirect pathway from vegetation variables (used as deforestation proxies) to next-hour precipitation via atmospheric intermediaries, consistent with the conceptual pathways shown in panel (a).}
    \label{fig:DAG}
\end{figure}

To determine whether the model learns merely predictive correlative patterns or internalises underlying mechanistic relationships between vegetation, atmospheric state, and precipitation, we probe its learned representations using targeted input ablations. Inputs are grouped into vegetation state (V), atmospheric intermediaries (A; surface fluxes and thermodynamics), and precipitation memory from the previous hour (P). If vegetation influences rainfall primarily through its modulation of atmospheric conditions, then removing vegetation information together with precipitation memory should disrupt forecast fidelity more strongly than removing precipitation memory alone. By contrast, if vegetation exerted a direct and instantaneous influence on rainfall beyond atmospheric state, removing vegetation inputs would substantially alter model behaviour even when atmospheric variables are retained. The observed pattern of responses across these ablations therefore distinguishes direct vegetation effects from indirect atmospheric mediation, revealing how vegetation information is organised within the learned dependency structure (Figure \ref{fig:DAG}; Table \ref{tab:ablation_results}). 

Removing vegetation inputs while retaining atmospheric variables and precipitation memory (A, P) leads to only a modest reduction in predictive performance relative to the full model (A, V, P), indicating that vegetation does not exert a strong instantaneous influence on next-hour precipitation when atmospheric state is already known. This is consistent with the absence of a dominant direct vegetation-precipitation pathway at hourly timescales ($V \centernot \rightarrow P$).

Removing the autoregressive precipitation signal (A, V) causes a substantial drop in performance, confirming that past precipitation is a key driver of the model's future rainfall prediction ($P_{t-1} \rightarrow P_t$). Most notably, removing both vegetation and precipitation inputs (A only) results in a larger performance decline than removing precipitation alone (A, V). This implies that vegetation provides complementary predictive information when precipitation memory is absent, through its effect on the intermediate atmospheric state ($V \rightarrow A \rightarrow P$). 

Together, these results are consistent with the model encoding vegetation effects on rainfall indirectly via atmospheric conditions, consistent with a front-door–type mediation structure \cite{Pearl2009} at the level of learned representations. This interpretation is further supported by a constraint-based causal graph analysis (PC algorithm), which recovers directional dependencies from vegetation variables to atmospheric state and onward to precipitation (Figure \ref{fig:DAG}; Supplementary Figure S3). We emphasise that these analyses do not identify causal effects in the physical system, but test whether the model’s learned dependency structure is consistent with a mechanistically mediated vegetation-atmosphere-precipitation pathway rather than shortcut correlations.

\FloatBarrier

\begin{figure}[htbp!]
    \centering
    \includegraphics[width=0.9\linewidth]{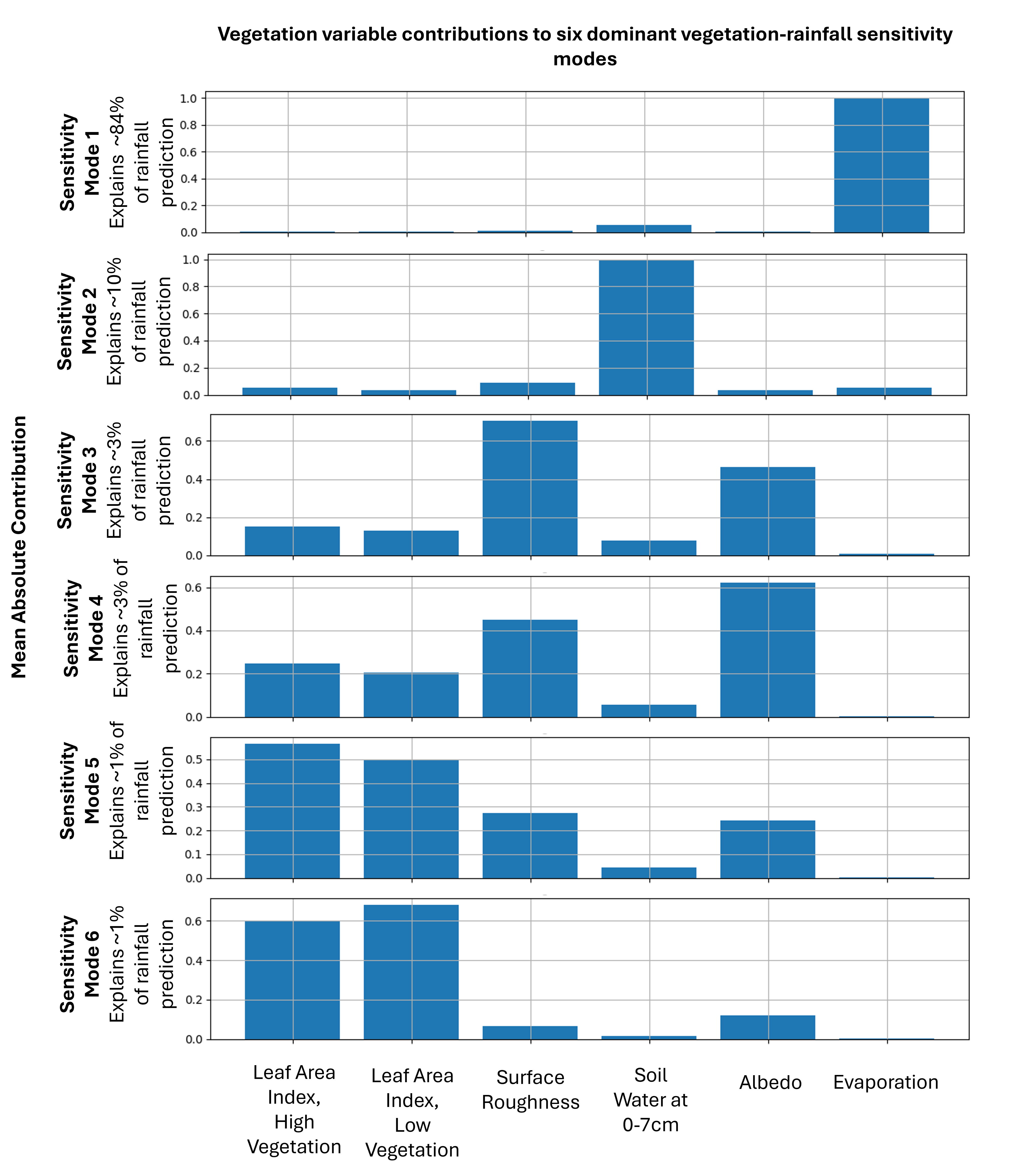}
    \caption{A series of bar charts showing the relative contributions of each vegetation variable to the six dominant vegetation-rainfall sensitivity modes identified by singular value decomposition (SVD) of the model’s vegetation-precipitation gradients. Modes are ordered by decreasing importance, with Mode 1 representing the dominant sensitivity pattern and Mode 6 the weakest. The percentages indicate the proportion of total variance in the model’s rainfall prediction explained by each mode, providing a quantitative measure of their relative influence on precipitation prediction. Contributions within each mode reflect the mean absolute loading of each vegetation variable. Variables with more immediate and direct influence on rainfall, such as evaporation, dominate the leading mode, while variables associated with slower land-surface feedbacks, such as LAI, contribute primarily to later modes. Although LAI explains a smaller fraction of short-term rainfall variance, the model remains responsive to LAI perturbations, indicating sensitivity to longer-timescale vegetation-atmosphere interactions (see Section 2.2.1).}
    \label{fig:gradients}
\end{figure}

To further assess whether the model internalises physically consistent vegetation-rainfall relationships, we examine the structure of the gradients learned during training. In this context, gradients quantify the local sensitivity of predicted precipitation to small perturbations in vegetation variables at preceding time steps, providing a physically interpretable measure of how changes in vegetation state are transmitted through the model’s representation to influence rainfall. By analysing coherent gradient patterns across space and time, we identify a small number of dominant influence axes that capture the principal ways in which vegetation variables jointly shape rainfall predictions (Figure \ref{fig:gradients}). These axes are ordered by the fraction of rainfall variance they explain, revealing a clear hierarchy of vegetation controls by timescale. Variables associated with rapid surface-atmosphere exchange, particularly evaporation, dominate the leading axes explaining 84\% of rainfall predictions, indicating that the model assigns them the greatest immediate influence on rainfall. Structural vegetation properties such as LAI emerge in later axes, consistent with their more indirect and temporally delayed role in modulating moisture recycling and boundary-layer development. This temporal separation implies that the model implicitly ranks vegetation variables according to the timescale of their hydrometeorological impact\cite{smith2023}: fast-acting drivers are prioritised early, while slower canopy-mediated effects appear later. Together, these results indicate that the model has internalised a delayed vegetation-rainfall response structure aligned with ecohydrological theory and consistent with its gated temporal memory.

\subsubsection{Deforestation Nonlinearly Skews Rainfall Distributions Toward Lower Intensities, with Asymmetric Responses Under Forest Regrowth}

To assess how rainfall responds to vegetation change, we compare predicted rainfall under counterfactual deforestation and regrowth scenarios to a control case with no perturbation. Vegetation loss is imposed in ecologically relevant hotspot regions, and resulting shifts in the full rainfall intensity distribution are analysed. Across all scenarios, we observe that rainfall distributions change significantly, with modest, spatially localised vegetation degradation nonlinearly altering short-term precipitation dynamics.

\begin{table}
\centering
\caption{Mean, maximum, and standard deviation of hourly precipitation prediction changes under different deforestation scenarios. All values are reported to four decimal places.}
\label{tab:deforestation_scenarios_defined}
\begin{tabular}{p{4.4cm}ccc}
\toprule
\textbf{Scenario} & \textbf{Mean Change (mm)} & \textbf{Max Change (mm)} & \textbf{Std. Dev. (mm)} \\
\midrule
Scenario 1: Heavy diffuse deforestation\\(-1.51\% per year) 
& 0.0043 & 11.7592 & 0.0216 \\

Scenario 2: Heavy local deforestation\\(-1.00\% per year) 
& 0.0035 & 9.4567 & 0.0174 \\

Scenario 3: Moderate deforestation\\(-0.38\% per year) 
& 0.0007 & 4.7214 & 0.0044 \\

Scenario 4: Light deforestation\\(-0.27\% per year) 
& 0.0005 & 3.5338 & 0.0034 \\

Scenario 5: Slow forest regrowth\\(+0.10\% per year) 
& 0.0015 & 0.8336 & 0.0015 \\
\bottomrule
\end{tabular}
\end{table}

\begin{figure}[htbp!]
    \centering
    \includegraphics[width=1\linewidth]{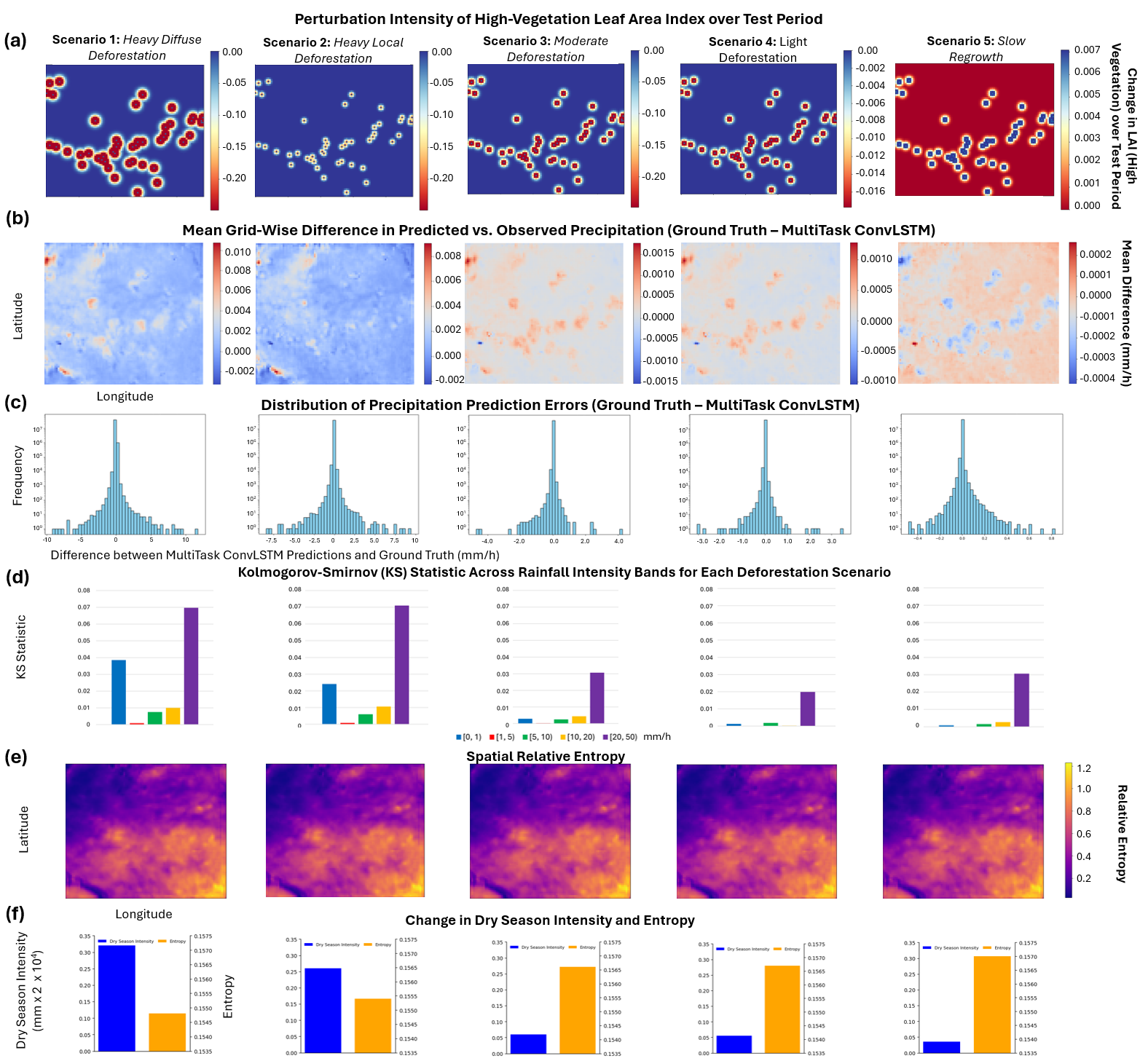}
    \caption{
    \textbf{a,} Spatial pattern of vegetation perturbations in the counterfactual experiments. The map shows the one-year accumulated reduction in high-vegetation LAI derived from a Gaussian decay centered on deforestation hotspots. Colour indicates perturbation magnitude; the same scheme was applied to all vegetation variables (see Methods).
    \textbf{b,} Grid-wise average difference in hourly precipitation (mm/h) between the control (no deforestation) scenario and each counterfactual deforestation scenario. Positive values indicate reduced rainfall under deforestation, and negative values indicate increased rainfall. Responses are spatially heterogeneous but consistently concentrated in the north-western Amazon, with diffuse changes propagating beyond direct deforestation hotspots. \textbf{d,} Distribution of grid-wise precipitation differences for each scenario. The histograms approximate skewed normal distributions with long tails in the direction of dominant rainfall change, indicating that while most regions experience small shifts, a minority exhibit pronounced precipitation responses. \textbf{d,} Kolmogorov-Smirnov (KS) statistics across rainfall intensity bins for each scenario (see Methods for details). Scenario 2 yields the strongest divergence in both moderate and heavy rainfall ranges. 
    \textbf{e, } Spatial relative entropy of precipitation distributions under each counterfactual deforestation scenario compared to the control (no deforestation). The increase in entropy is most pronounced under Scenario~5 (reforestation).
    \textbf{f,} Comparison of dry season intensity and overall precipitation entropy across scenarios. Severe deforestation (Scenarios~1 and 2) leads to substantial increases in dry season intensity, while Scenarios~3-4 (slower deforestation) result in only moderate increases. Scenario~5 yields a slight reduction in dry season intensity. The figures shows consistent shift towards spatially ordered rainfall under deforestation.}
    \label{fig:scenarios}
\end{figure}

As seen in Figures~\ref{fig:scenarios} and Table~\ref{tab:deforestation_scenarios_defined}, Scenario~1, simulating the strongest vegetation loss (approximately 1.51\% annual forest cover reduction) over a broad spatial extent (hotspot zones $\approx$ 50km wide), produced the largest response: a maximum deviation of 11.76~mm/h and a mean of 0.0043~mm/h. These shifts indicate that diffuse canopy thinning across widespread hotspots can cause widespread but moderate hydrological disruption. When the same intensity of deforestation was applied to a more spatially concentrated area (Scenario~2, hotspot zones $\approx$ 5km wide), approximately 1.00\% annual cover loss), the mean rainfall change slightly decreased by 18\% compared to Scenario~1, with a maximum deviation of 9.46~mm/h. However, the proportion of moderate rainfall changes (5--10~mm/h) increased, as seen in Figure~\ref{fig:scenarios} \textbf{(c)}. The Kolmogorov-Smirnov (KS) statistic, shown in Figure~\ref{fig:scenarios} \textbf{(d)}, revealed that Scenario~2 produced a slightly greater maximum divergence in the cumulative rainfall distribution (KS = 0.071) than Scenario~1 (KS = 0.068). This means that at some rainfall threshold, the probability of occurrence shifted by up to 7.1 percentage points under the perturbation. Conversely, light rainfall events (below 1~mm/h) were more affected in Scenario~1, with a distributional divergence of 0.0385 compared to 0.0241 in Scenario~2. The remaining distribution bands remained largely stable across both scenarios. These findings correspond to spatially concentrated deforestation intensifying moderate rainfall bands, while more widespread but diffuse canopy loss primarily affects lighter rainfall. 

Scenarios~3 and~4 correspond to typical contemporary deforestation rates in the Amazon (0.38\% and 0.27\% forest loss per year, hotspot zones approximately 20km wide). Although the mean hourly rainfall changes are small (0.0007 and 0.0005~mm/h), the rainfall distribution departs systematically from the control case (Figure~\ref{fig:scenarios}\textbf{d}). The strongest divergence occurs in the upper tail (20--50~mm/h), indicating that moderate annual deforestation disproportionately suppresses the frequency of intense rainfall events. Weaker but non-negligible divergence is also observed in intermediate rainfall bands (10--20~mm/h) and at very low intensities (0--1~mm/h), suggesting that vegetation loss perturbs not only extremes but the overall shape of the rainfall distribution. These secondary shifts imply a redistribution of rainfall away from intense events toward lighter or fragmented precipitation regimes. Overall, this pattern corresponds to vegetation loss altering rainfall in a nonlinear manner, with the upper tail being most sensitive even under modest deforestation rates.

Scenario~5, which simulated mild reforestation (0.1\% forest cover gain per year), yielded the smallest hydrological impact: a maximum rainfall difference of 0.83~mm/h but with a mean change of 0.0015~mm/h, higher than both Scenario~3 and Scenario~4. KS values were lowest across all rainfall bins in this scenario. However, a non-zero KS statistic of 0.0306 was observed in the extreme rainfall bin, similar in magnitude to the value from Scenario~3, despite Scenario~3 involving nearly four times the annual vegetation change. This asymmetry implies that reforestation does not simply reverse the effects of deforestation; instead, it has a disproportionate effect on extreme rainfall events without equivalently shifting other rainfall bands, highlighting the directionally sensitive nature of vegetation-precipitation feedbacks learned by the model, where the reintroduction of vegetation fails to fully restore evapotranspiration rates, convective triggers, or soil-atmosphere coupling lost during deforestation. Such hysteresis effects, where ecosystem responses depend on historical trajectories, are well-documented in coupled land-atmosphere systems and highlight the risk of crossing ecological tipping points beyond which recovery is incomplete or delayed \cite{baudenaEffectsLanduseChange2021}.

Spatially, precipitation changes were concentrated in the Western Amazon and Andean foothills (Figure~\ref{fig:scenarios} \textbf{(b)}). Under strong deforestation (Scenario~1), rainfall reductions spanned the northwest and western periphery and diffuse areas around deforestation hotspots, while increases were mostly confined offshore. The Amazon interior remained largely unchanged, appearing to function as a hydrological buffer. Histogram analysis showed an overall negative skew with a heavier upper tail, indicating more frequent under-prediction with occasional large spikes in over-prediction. In Scenario~2, spatial disruptions were more confined but retained comparable KS statistic extremity, reaffirming that smaller, sharper deforestation events can provoke abrupt shifts within limited areas. Scenarios~3 and~4 sustained these spatial patterns, with rainfall reductions clustering in dry western and northern edges and in diffuse regions around deforestation hotspots. The Amazon interior remained largely invariant, and histograms maintained a slight negative skew, with Scenario~3 again displaying stronger deviations in line with its higher perturbation rate. In contrast, Scenario~5 produced gentle increases in precipitation across hotspot zones and particularly near the Andes, while modest reductions appeared in structurally sensitive inland zones, including around Lake Titicaca. The resulting histogram showed a mild positive skew, corresponding to reforestation modestly improves rainfall predictions. Together, these findings correspond with minor canopy alterations near topographically or climatologically sensitive zones propagating structured shifts in predicted rainfall regimes.

\subsection{Vegetation Loss Intensifies Dry Seasons and Decreases Rainfall Entropy}

Monthly aggregates of the hourly predictions are summarised with Relative Entropy (RE), Dry Season Intensity (DSI), via Penman-Monteith \cite{Penman1948}, and PAF to capture seasonality, hydrological stress, and spatial extent (see Methods). 
As shown in Figure~\ref{fig:scenarios} \textbf{(e), (f)}, deforestation scenarios consistently increased entropy across affected regions in comparison to the control scenario, with the Southeastern Amazon showing the most marked rise. This suggests that vegetation loss introduces disorder into the model’s prediction of local rainfall dynamics, especially in areas with already-high uncertainty. Intriguingly, reforestation (Scenario~5) yielded the largest overall increase in entropy (Figure~\ref{fig:scenarios} \textbf{(f)}, corresponding to the potential for vegetation recovery to temporarily destabilise the rainfall distribution, possibly due to delayed feedbacks in surface moisture and energy fluxes. Analysis of dry season characteristics further reinforced these findings (Figure~\ref{fig:scenarios} \textbf{(f)}). Dry season intensity increased markedly under strong deforestation: in Scenarios~1 and~2, it reached 6600~mm and 5200~mm, respectively, indicating elevated water deficits during periods of extensive forest loss. In contrast, Scenario~3 saw a substantial drop to 1200~mm, with a further slight reduction to 1100~mm in Scenario~4. Scenario~5, representing reforestation, exhibited an almost negligible dry season intensity.
The sharp contrast between Scenarios~1 and~2 and the remaining scenarios corresponds to a nonlinear relationship between deforestation magnitude and drought severity. Notably, the difference in dry season intensity between the moderate deforestation and reforestation scenarios was considerably smaller than between the two high-deforestation cases. This corresponds to the fact that light to moderate deforestation may not drastically intensify dry seasons, while even small-scale reforestation exerts a drought-mitigating effect on cumulative hourly rainfall.
Overall, these results show that the model captures rainfall sensitivity to vegetation change in a manner consistent with known biophysical mechanisms. These results correspond to the fact that reduced forest cover intensifies and destabilises rainfall, particularly in the Western Amazon and dry frontier zones, while reforestation moderates these effects, disproportionately impacting extreme rainfall events.

\subsubsection{Rainfall Sensitivity Varies Strongly Across Vegetation Properties and Regions}

To quantify how strongly rainfall responds to different vegetation properties, we identify the smallest changes in each vegetation variable required to produce a measurable shift in predicted precipitation. To do this, we apply Projected Gradient Descent (PGD) that incrementally adjusts vegetation variables within physically plausible bounds, stopping once rainfall intensity or the spatial extent of rainfall changes beyond a predefined threshold of PAF (see Methods). By repeating this process across the domain, and with perturbations restricted to deforestation hotspots or unrestricted, we map where rainfall is most sensitive to vegetation change and which vegetation variables exert the strongest control.

\begin{figure}[htbp!]
    \centering
    \includegraphics[width=1\linewidth]{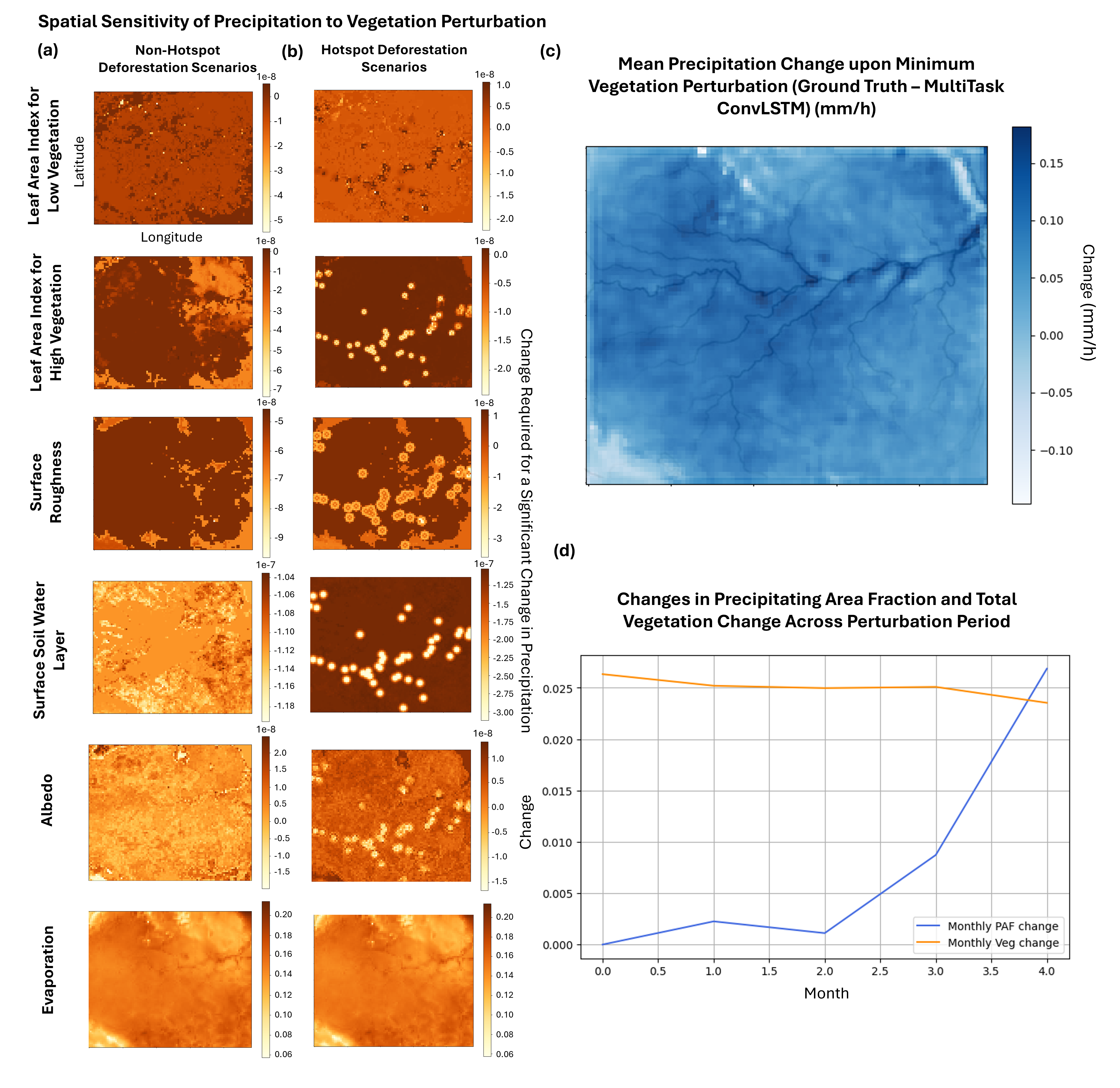}
    \caption{
    \textbf{a,} PGD-derived vegetation sensitivity maps showing the mean perturbation required to elicit a significant change in predicted precipitation across six vegetation variables (LAI-low, LAI-high, surface roughness, soil water, albedo, and evaporation). These perturbations are restricted to non-hotspot zones. \textbf{b,} Equivalent vegetation sensitivity maps with perturbations constrained to historical deforestation hotspots. White regions indicate high sensitivity (i.e., minimal change needed to affect precipitation). Vegetation perturbations are shown in physical units. For reference, climatological ranges over the study region are: high-vegetation LAI (0--7.09), low-vegetation LAI (0--3.61), surface roughness (2.4×10$^{-5}$--1.98), soil water (0--0.59), albedo (0.011--0.133), and evaporation (-1.0×10$^{-3}$--5.6×10$^{-5}$). \textbf{c,} Mean change in predicted hourly precipitation across the Amazon resulting from minimum vegetation perturbations (aggregated across variables). Positive values (darker regions) indicate precipitation loss. Major river channels (Amazon and tributaries) are overlaid to aid interpretation. \textbf{d,} Monthly evolution of PAF under a hypothetical worst-case vegetation perturbation scenario. Vegetation variables are incrementally perturbed in the locally most rainfall-sensitive directions identified in the anlaysis, representing a stress-test. While vegetation perturbations remain relatively stable across months, the model's PAF sensitivity increases markedly after month 2, with a sharp rise after month 3. This suggests that the model becomes progressively more responsive to vegetation changes over time, with delayed but amplified hydrological impacts. The PGD procedure, stopped only upon reaching a target precipitation or PAF shift, thus reveals a latent, cumulative sensitivity in the model’s rainfall dynamics that emerges after sustained vegetation perturbation.}
    \label{fig:sensitivies_and_PAF}
\end{figure}

\FloatBarrier

In Figure \ref{fig:sensitivies_and_PAF} \textbf{(a, b)} we see that LAI exhibited the highest average change under PGD, with non-localised effects appearing in peripheral rainforest zones. This suggests that forest canopy structures influence the model's predicted rainfall both locally and indirectly, possibly through delayed transpiration feedbacks. Surface roughness elicited rainfall sensitivity in ring-shaped bands around deforestation cores, with the highest rainfall responses observed in transitional zones. This supports the idea that rainfall reacts more strongly to spatial gradients (edge effects) than to absolute roughness values. Soil water content responded most predictably and locally to vegetation perturbation. The model's predictions of rainfall sensitivity to soil moisture in the early months is consistent with its known role as a fast responder to vegetation loss via reduced infiltration \cite{sanchezmartinez2025}. Albedo had context-dependent impacts \cite{doughty2012}. Sensitivity to albedo was concentrated over aquatic regions under hotspot-constrained perturbations and in arid zones under uniform perturbation, implying that the model has implicitly learned conditional relationships between surface type and rainfall response. Evaporation, as expected, had the most direct and intense effect on precipitation, with its influence extending well beyond hotspot regions and persisting across all PGD directions.

\subsubsection{Vegetation-Rainfall Tipping Point Emerges Under Three Months of Sustained Vegetation Perturbation}

To test whether rainfall responds smoothly or exhibits threshold-like behaviour, we examine the temporal evolution of rainfall sensitivity under sustained vegetation perturbations applied in the most rainfall-sensitive directions. Vegetation variables are incrementally perturbed in a consistent, worst-case configuration identified by the sensitivity analysis. By tracking how the minimum perturbation required to alter rainfall, together with changes in rainfall intensity and PAF, evolves over successive months, we assess whether rainfall responses accumulate gradually or emerge abruptly after sustained forcing (see Methods).

The vegetation sensitivity experiments show that the model’s predicted rainfall response to sustained vegetation perturbations is nonlinear and cumulative.  PAF remains largely unchanged during early perturbation stages but increases markedly after Month~3, suggesting that sustained vegetation forcing leads to an abrupt contraction in the spatial extent of predicted rainfall (Figure~\ref{fig:sensitivies_and_PAF}). This behaviour reflects the model’s learned response to persistent vegetation stress. These results align with ecohydrological theory, which predicts abrupt shifts in rainfall once biospheric limits are exceeded \cite{flores2024}. They also reinforce the notion that tipping points in vegetation-rainfall coupling are regionally contingent, particularly in the Western and Northeastern Amazon, where many changes were spatially concentrated, as seen in Figure~\ref{fig:scenarios} \textbf{(c)}.

\section{Discussion and Conclusion}

Our results correspond to the fact that deforestation leads to a net reduction in rainfall across the Amazon Basin, with pronounced spatial heterogeneity such that drying outweighs compensatory increases at the basin scale. The north-western Amazon emerges as the most sensitive region, as predicted by our model, with precipitation reductions reaching up to 11.76~mm~h$^{-1}$ and $\approx$ 3~mm~month$^{-1}$ under high deforestation. Localised buffering and partial reversals occur near major aquatic systems, where sustained evaporation maintains high latent heat fluxes that partially offset reduced transpiration. Evaporation from rivers and lakes sustains boundary-layer humidity and convection, mitigating rainfall losses in adjacent deforested areas, consistent with process-based studies \cite{lejeune2014}.

These spatial responses align with previous findings identifying the north-western Amazon as strongly dependent on moisture recycling, exhibiting resistance to moderate rainfall decline but heightened vulnerability once critical thresholds are exceeded \cite{nian2024}. Rainfall impacts extend beyond deforested regions into downwind areas, reflecting altered moisture transport and mesoscale circulation that propagate anomalies across the basin \cite{smith2023}. In our simulations, these redistributions decrease rainfall entropy by shifting rainfall away from the north-western Amazon while enhancing precipitation in parts of the south and north-eastern coast, with the largest reductions occurring downwind of deforestation hotspots, consistent with earlier modelling studies \cite{qin2025, medvigy2011}. Under deforestation, increased rainfall heterogeneity is accompanied by systematic dry-season intensification, consistent with evidence that climate variability over tropical land is increasingly dominated by seasonal asymmetry, with dry seasons becoming drier and more persistent \cite{konapala2020}. Reforestation scenarios produce the largest increases in rainfall entropy relative to the no-deforestation baseline; however this may be attributed to the model being trained on observational data dominated by progressive forest loss, perturbations toward regrowth may lie outside the learned dynamical regime, leading to broader rainfall distributions and increased predictive uncertainty.

When aggregated over time, the magnitude of rainfall change predicted by the model is broadly consistent with existing observational and modelling evidence \cite{esquivel-muelbertCompositionalResponseAmazon2019, qin2025, 29}. Across deforestation scenarios, we observe systematic dry-season rainfall declines that intensify with increasing forest loss, alongside localised increases under mild regrowth. For example, dry-season rainfall decreases reach $-3.14$~mm~month$^{-1}$ at 1.5\% forest loss, $-2.56$~mm~month$^{-1}$ at 1.0\%, and $-0.36$~mm~month$^{-1}$ under moderate deforestation (0.27--0.38\%), with a local increase of $+1.1$~mm~month$^{-1}$ under 0.1\% reforestation. These spatially coherent responses are consistent with regional drying patterns associated with altered moisture recycling and mesoscale circulation, as reported in process-based studies \cite{qin2025}. Similarly, under business-as-usual (moderate) deforestation scenarios, we find dry-season rainfall reductions of 0.19--0.21\% after a single year, comparable in scale to multi-decadal declines inferred in longer-term projections \cite{29}. While direct extrapolation across timescales should be treated cautiously, the agreement in magnitude indicates that short-timescale vegetation-atmosphere sensitivities captured by the model plausibly accumulate into longer-term rainfall change. Resolving these effects at hourly resolution enables detection of early redistribution and nonlinear organisation of rainfall that would be obscured by analyses based solely on monthly or seasonal averages, allowing short-term precursors to longer-term shifts to be identified.

Rainfall sensitivity, as predicted by the model, is strongest at the peripheries of deforestation hotspots, where albedo, surface roughness, and high-vegetation LAI exert the dominant influence, while low-vegetation LAI produces an opposing effect. This spatial structure reflects a separation between circulation-driven responses associated with sharp surface contrasts and evapotranspiration-driven local controls, consistent with previous studies \cite{qin2025, konapala2020, 29}. Increased seasonal variability in rainfall under deforestation has been linked to land-surface radiative effects, particularly changes in surface albedo that modify net radiation and boundary-layer stability \cite{18}, consistent with our finding that rainfall is especially sensitive to albedo perturbations around hotspot edges. At these boundaries, abrupt contrasts in albedo and roughness generate strong surface heterogeneities that promote small-scale circulations and focused low-level convergence, amplifying rainfall sensitivity. Away from hotspot edges, where surface contrasts weaken, rainfall sensitivity is instead governed by vegetation-driven moisture supply. In these regions, LAI regulates canopy transpiration and boundary-layer humidity, shaping the thermodynamic conditions for convection: high LAI supports moisture recycling, whereas reduced LAI limits atmospheric moisture availability and increases susceptibility to rainfall suppression. While these mechanisms are well established in process-based modelling, here they emerge from an interpretable neural framework with explicit temporal ordering at hourly scales. Sensitivity decomposition identifies evaporation as the dominant short-timescale control on rainfall, followed by soil moisture, surface roughness, albedo, and LAI. This hierarchy is consistent with ecohydrological theory and, to our knowledge, has not previously been extracted from neural rainfall models.
Beyond changes in mean rainfall, sustained vegetation perturbations nonlinearly redistribute the rainfall intensity spectrum. Under moderate deforestation, heavy rainfall (20–50~mm~h$^{-1}$) predictions declines by up to 7\%, while light rainfall (0–1~mm~h$^{-1}$) predictions increases by up to 4\%, indicating a structural reorganisation of rainfall. Reforestation produces an asymmetric response, disproportionately enhancing extreme rainfall predictions while inducing comparatively small changes in mean precipitation prediction, consistent with short-timescale hysteresis in vegetation-rainfall coupling \cite{staal2020}. Such distributional responses have not, to our knowledge, been systematically quantified at hourly timescales in deforestation-rainfall studies.

A growing body of work suggests that Amazonian rainfall may exhibit tipping behaviour under sustained deforestation, with proposed thresholds ranging from $\sim$20\% forest loss to basin-scale transitions when 30–50\% of the forest is degraded, typically inferred from seasonal or climatological aggregates \cite{2zemp2017, boers2017, staal2020, nian2024}. By resolving rainfall responses at hourly timescales, our analysis provides a complementary perspective on how nonlinear behaviour may emerge. We find that when vegetation variables are persistently perturbed in the most sensitive regions and directions, the precipitating area fraction remains stable before undergoing an abrupt decline after approximately 2–3 months, as predicted by the model. This response corresponds to stress-induced reorganisation of vegetation–atmosphere coupling, demonstrated here through a short-timescale stress test rather than long-term land-cover change. Importantly, this behaviour should not be interpreted as evidence of a formal climatic tipping point. It indicates that rainfall predictions can reorganise rapidly once critical stress is sustained, potentially serving as an early manifestation of threshold-like behaviour.

We note that model predictive skill was sufficient to support meaningful attribution. Previous deep-learning nowcasting systems achieve high skill at low rainfall thresholds but degrade rapidly for heavier precipitation. DGMR~\cite{Ravuri2021} reports CSI values of approximately 0.5 at 1~mm~h$^{-1}$, falling below 0.2 at 4–8~mm~h$^{-1}$, while MetNet-2~\cite{espeholt2022} reports CSI values of 0.45 at 2~mm~h$^{-1}$, 0.20 at 8~mm~h$^{-1}$, and 0.15 at 20~mm~h$^{-1}$. Our MultiTask ConvLSTM attains comparable skill at low intensities (CSI = 0.49 at 1~mm~h$^{-1}$) and improved performance at higher thresholds (CSI = 0.24, 0.30, and 0.39 at 2, 8, and 20~mm~h$^{-1}$, respectively), with corresponding F1 scores of 0.69, 0.62, and 0.45. Skill remains lowest at intermediate intensities, consistent with the intermittent nature of convective rainfall. While a trade-off between predictive accuracy and interpretability is a recognised limitation in machine learning \cite{dziugaite2020, lovo2024}, we achieve interpretability of the model through targeted ablation experiments, gradient-based sensitivity analysis, and weight decomposition, allowing verification that learned dependency structures are consistent with established land-atmosphere mechanisms. The model functions as an interpretable experimental tool, enabling vegetation-driven rainfall sensitivities to be analysed at convective timescales not accessible in most existing frameworks.

\subsection{Limitations}

While the MultiTask ConvLSTM demonstrated strong predictive accuracy and interpretability, several limitations constrain the generalisability and strength of our conclusions. The model was trained on only 1.5 years of hourly data, limiting exposure to interannual variability and long-timescale land-atmosphere interactions. As a result, slow-varying climatic drivers and delayed feedbacks may be underrepresented in the learned dynamics. Although performance remained robust across two independent test periods, including during extreme drought conditions, relationships learned over this limited temporal window may not fully generalise to longer-term climate regimes. In addition, ERA5-Land reanalysis integrates heterogeneous observational sources within a constrained assimilation framework, which may underrepresent uncertainty in remote or densely vegetated regions such as the central Amazon. These uncertainties propagate into the model and may affect inference fidelity.
The spatially fixed grid and regional training domain further restricted the model’s capacity to account for external climatic drivers. Teleconnections such as the El Niño-Southern Oscillation Cycle or Atlantic Multidecadal Oscillation Cycle were excluded, limiting applicability to broader climate shifts. Moreover, the unified ConvLSTM processing of vegetation and atmospheric inputs introduces spatial and temporal smoothing, reducing sensitivity to sharp gradients or regime transitions. This is consistent with the observed degradation in performance for intermediate rainfall intensities, where precipitation is fragmented, short-lived, and highly sensitive to small-scale triggering processes. As a result, the model performs best at the extremes of the rainfall distribution and less reliably within the transitional “belly” regime.

Our counterfactual scenarios, gradient-based analyses, and interventional ablations should be interpreted cautiously. These experiments probe vegetation-rainfall coupling within the model but simulate hypotheses rather than physical climate futures, as all perturbations operate through internal learned representations. Interventional ablations may introduce both visible and hidden confounding, for example through correlations between vegetation variables such as LAI and soil moisture, or through latent features embedded within the network, which can create spurious associations or mask true causal pathways. Deep neural networks are known to exploit shortcut correlations and embedded priors \cite{Lapuschkin2019, Zhou2022}, and Earth system dynamics themselves can generate non-identifiable or spurious causal structures \cite{Runge2019, Pearl2009}. Weight decomposition, visualisation, and saliency analyses (Supplementary Notes S2-S3) therefore illustrate model sensitivities but cannot guarantee causal correctness, and may reflect Clever Hans artefacts \cite{Lapuschkin2019}. To mitigate this, we combined PGD perturbations with spatial fidelity metrics (CSI and CRPS) to distinguish representational saliency from predictive impact; for example, variables that appear salient but produce no effect under PGD would indicate representational confounding. We did not observe such inconsistencies, nor did causal graph discovery using the PC algorithm identify explicit confounders linking vegetation and precipitation. However, the absence of detected confounders should not be interpreted as evidence that confounding is negligible, but rather as a consequence of the limited temporal span of the training data, which restricts the ability to resolve slow-varying or externally driven confounders. Consequently, although ablation and sensitivity analyses consistently support an indirect vegetation-atmosphere-precipitation pathway within the model, these results do not satisfy the formal conditions of the front-door criterion. They only indicate that the model has internalised a mechanistic structure consistent with established ecohydrological mechanisms. More generally, Earth system modelling is fundamentally underdetermined, and multiple causal structures can explain observed time series equally well \cite{Runge2019}. Robust causal inference would require explicit structural priors, longer observational records, or experimental validation, which remain beyond the scope of this study. Accordingly, our causal analyses should be viewed as hypothesis-generating rather than definitive ecological or policy conclusions.

Finally, the sensitivity analysis framework assumed independence among vegetation variables, neglecting their ecological coevolution, for example, between LAI, albedo, and roughness, which may bias estimated sensitivities. More sophisticated joint perturbations or structural causal models would be required to disentangle these interactions.

Despite these limitations, the model consistently identified plausible spatial and temporal precipitation responses to vegetation loss, supporting machine learning as a valuable tool for generating ecohydrological hypotheses. Future work should extend training to longer climate records, incorporate global-scale drivers, and validate inferred sensitivities with satellite or field observations.

\subsection{Policy Implications}

Our results indicate that Amazonian rainfall organisation can exhibit tipping-like responses to sustained vegetation degradation on timescales of months, as predicted by our model, with direct implications for land-use monitoring and policy. These findings emphasise the need to move beyond aggregate deforestation metrics towards short-term, spatially targeted assessment of rainfall sensitivity to specific land-surface characteristics.

We show that rainfall sensitivity is observed to depend on distinct vegetation and surface properties, with particularly strong responses to changes in radiative balance, such as surface albedo, around deforestation boundaries. As deforestation progresses, the set of land-surface characteristics to which rainfall is observed to be the most sensitive evolves, reflecting the weakening of biological moisture recycling and increasing dominance of surface energy controls. From a policy perspective, understanding how rainfall sensitivity shifts across vegetation characteristics can inform reforestation and land-management strategies that target canopy recovery, and also the surface properties most relevant for stabilising rainfall. Interventions that restore or manage vegetation in ways aligned with these sensitivities may help promote smoother rainfall recovery and reduce the risk of asymmetric or unpredictable hydrological responses during regrowth.

More broadly, these results demonstrate how interpretable, data-driven models can complement process-based climate models by identifying short-term vulnerability pathways and early-warning signals in the land-atmosphere system. Translating such sensitivity-informed diagnostics into policy will require integration with longer observational records, mechanistic modelling, and empirical validation, but it offers a pathway toward more responsive and risk-aware land-use governance.

\section{Materials and Methods}

\subsection{Dataset and Preprocessing}

We used ERA5-Land hourly reanalysis data on single levels from the Copernicus Climate Data Store\footnote{\url{https://cds.climate.copernicus.eu/cdsapp#!/dataset/reanalysis-era5-single-levels}}, covering the period from 1 January 2021 to 30 September 2024 (32,832 hours, 257,961,024 individual grid points). Data were extracted over the central Amazon Basin (latitude: 5°N to 15°S, longitude: 75°W to 50°W) at approximately 0.25° spatial resolution, yielding an $81 \times 97$ grid. The precipitation data exhibited a highly imbalanced distribution, with most values concentrated around 0 mm/h and extreme values reaching up to 45 mm/h. The training, validation, and test sets maintained similar histogram distributions, although minor spatial discrepancies were observed, most notably, reduced rainfall in Venezuelan cities in the northern subregion of the validation set. The mean precipitation value across the dataset was 0.24 mm/h, with a standard deviation of 0.71 mm/h.

Thirteen variables were selected based on their relevance to land-atmosphere coupling and deforestation dynamics. Surface Fluxes ($F_{t-1}$) included: surface latent heat flux, evaporation. Large-Scale and Near-Surface Thermodynamic Variables ($T_{t-1}$) included: 2m air temperature, 2m dew point temperature, 10m u- and v-wind components, and total column rainwater, all of which influence convection, humidity, and moisture transport. These latter two groups are generalised as atmospheric mediators ($A_{t-1}$) within this study. Vegetation and Land Surface Variables ($V_{t-1}$) related inputs comprised high and low LAI, surface roughness, surface albedo, and top-layer soil moisture (0-7 cm depth), serving as proxies for forest cover and degradation~ \cite{14,15,16,17,18,19,20}. Precipitation at the previous hour also served as an input ($P_{t-1}$). The prediction target was hourly total precipitation ($P_{t}$) at each grid cell.

All variables were standardised using z-score or log normalisation depending on distribution skew, with statistics computed from the training set only. This was done to handle data imbalance without increasing computational complexity. At each timestep, the model received a tensor of shape $[(T=1), H, W, C]$, where $C=14$ corresponds to the input channels across spatial dimensions $H \times W$. An extra binary input channel was added to aid with precipitation classification.

\subsection{MultiTask ConvLSTM for Rainfall Prediction}
\subsubsection{Model Formulation}

The architecture consists of a three-layer ConvLSTM encoder (for more details please see Supplementary Text, Supplementary Note 1), followed by a Conv3D layer that refines the shared spatiotemporal features, and two task-specific output heads: (i) a multi-layer perceptron (MLP) regression head for rainfall intensity, and (ii) a Conv3D classification head for rainfall occurrence. The network simultaneously learns spatially coherent rainfall/no-rain patterns and intensity distributions. The classification head uses binary cross-entropy loss. The regression head uses mean squared error as its loss function. The two are combined to form a compound loss function that is backpropagated during training. Since rainfall prediction is strongly imbalanced, with nearly half of all grid-point hours containing no rain (0 mm/h), pure regression losses performed poorly in this setting, since the model can minimise error by overpredicting dry states and underestimating rare but important rainfall events. Our multi-task formation explicitly teaches the model to capture dry-to-wet transitions while still learning rainfall magnitude. Alternative imbalance-mitigation methods were tested, but this joint formulation proved most effective, particularly for light and extreme rainfall tails (for more details, please see Supplementary Text, Supplementary Note 2).

Given an input tensor $X_t \in \mathbb{R}^{(T=1) \times C \times H \times W}$, $C$ is the number of input channels , and $H \times W$ is the spatial resolution, the model outputs two predictions per grid cell:

\begin{itemize}
    \item $\hat{y}^{\text{reg}}_t \in \mathbb{R}^{H \times W}$: rainfall intensity (mm/h),
    \item $\hat{y}^{\text{cls}}_t \in [0,1]^{H \times W}$: probability that rainfall is absent.
\end{itemize}

\vspace{0.5mm}
The MultiTask ConvLSTM has two output heads for rainfall classification and regression formulated as follows:
\begin{align*}
\hat{y}^{\text{cls}}_t &= \text{Conv3D}(H_t) \\
\hat{y}^{\text{reg}}_t &= \text{MLP}(H_t)
\end{align*}

For the regression head, a multi-layer perceptron (MLP) maps the 64-channel latent representation to a single precipitation intensity per grid cell. This dimensionality reduction forces the model to compress high-level spatiotemporal features into a scalar hydrological output through flexible nonlinear mappings.

For the classification head, a second Conv3D layer is used instead of an MLP. Rainfall occurrence is spatially coherent, so convolution at the output stage enforces spatial consistency and suppresses isolated false positives in dry regions (for architecture details, please see Supplementary Text, Figure S1).

\vspace{0.5mm}
Rainfall predictions are masked at inference time to reduce false positives in dry zones:
\[
\hat{y}_{t,ij} =
\begin{cases}
0 & \text{if } \hat{y}^{\text{cls}}_{t,ij} > 0.6 \text{ and } \hat{y}^{\text{reg}}_{t,ij} < 0.1 \\
\hat{y}^{\text{reg}}_{t,ij} & \text{otherwise}
\end{cases}
\]

\vspace{0.5mm}
We defined rainfall as present when intensity $\geq 0.1$ mm/h:
\[
y^{\text{cls}}_{t,ij} = \mathbb{1}[y^{\text{reg}}_{t,ij} < 0.1]
\]

\vspace{0.5mm}
The total loss balances regression and classification objectives:
\[
\mathcal{L}_{\text{total}} = \frac{1}{2} \left( \mathcal{L}_{\text{reg}} + \mathcal{L}_{\text{cls}} \right)
\]
\begin{align*}
\mathcal{L}_{\text{reg}} &= \frac{1}{HW} \sum_{i,j} (y^{\text{reg}}_{t,ij} - \hat{y}^{\text{reg}}_{t,ij})^2 \\
\mathcal{L}_{\text{cls}} &= \frac{1}{HW} \sum_{i,j} \left[ y^{\text{cls}}_{t,ij} \log \hat{y}^{\text{cls}}_{t,ij} + (1 - y^{\text{cls}}_{t,ij}) \log (1 - \hat{y}^{\text{cls}}_{t,ij}) \right]
\end{align*}

\subsubsection{Hyperparameter Selection and Training Protocol}

The MultiTask ConvLSTM consists of a shared 3-layer encoder followed by a 3D convolutional layer and two task-specific heads: a Conv3D layer for rainfall occurrence classification and a 3-layer MLP for rainfall intensity regression. Training used a composite loss combining classification and regression terms (Section 4.2.1). Hyperparameters were tuned using a validation dataset spanning 2023 data. As preprocessing, log-normalisation of input rainfall with offset 1 and MSE as a loss function were selected as they yielded superior performance for low-intensity rainfall (see Supplementary Text, Supplementary Note 2). Final hidden dimensions (32, 64) and kernel size (3×3) provided the best balance of stability and capacity (see  Supplementary Text, Supplementary Note 2). Weight distributions across ConvLSTM, batch normalization, and output layers are provided in Supplementary Text, Figure S2, confirming well-regularized training. 

\subsubsection{Evaluation Protocol}

Model evaluation was conducted on two independent test periods: 24 May-31 December 2023 and 1 January-30 September 2024, the latter representing an anomalously dry season. The former was also used for counterfactual testing.

Metrics were selected by an extensive literature review on precipitation nowcasting \cite{duku2023, 21, xu2024, wang2021multi, liao2024short, CNN_convective, convlstmhybrid, transformer, espeholt2022, Sonderby2020, Ravuri2021}, chosen to capture a balance of error magnitude, forecast fidelity, temporal correlation, and spatial pattern accuracy.

For rainfall regression, we used Mean Squared Error (MSE) to quantify average prediction error, and Nash-Sutcliffe Efficiency (NSE) to assess forecast fidelity relative to observed variance. Pearson, Spearman, and Kendall correlations were used to capture linear, monotonic, and rank-based temporal agreement, respectively. The Continuous Ranked Probability Score (CRPS) evaluates probabilistic forecasts, asking whether predicted rainfall distributions are both accurate and sharp, which is critical in weather prediction. For rainfall occurrence classification, we used accuracy, F1 score to balance precision and recall, and ROC-AUC to assess threshold-independent separability. A threshold of 0.1~mm/h was used to binarise rainfall. The Critical Success Index (CSI) specifically measures how often rainfall was correctly detected when compared against misses and false alarms, and does so at different rainfall intensities, making it especially useful for assessing performance on light, moderate, and extreme events. Metrics were computed at hourly resolution and monthly aggregates, with regression scores averaged across all grid cells unless otherwise specified.

\subsection{Mechanistic and Gradient-Based Attribution}
\subsubsection{Mechanistic Ablation}

We aimed to test whether the MultiTask ConvLSTM has learned physically consistent, indirect mechanistic pathways linking vegetation to rainfall, such that perturbations to vegetation propagate through atmospheric mediators.

To examine how vegetation perturbations are represented within land-atmosphere coupling, we developed a mechanistic intercomparison framework linking four process groups: Vegetation ($V$: LAI, albedo, surface roughness, soil water), Surface Fluxes ($F$: latent heat flux, evaporation), Thermodynamics ($T$: temperature, dewpoint, rainwater, winds), and Precipitation ($P$). Precipitation memory from the previous hour ($P_{t-1}$) was included explicitly to capture persistence effects such as wetting/drying, convective cold pools, and canopy stress. Grouping was motivated by physical reasoning: (i) $V \leftrightarrow F$ captures the bidirectional exchange of water and energy via evapotranspiration and canopy conductance; (ii) $F \leftrightarrow T$ reflects surface-atmosphere coupling of heat and moisture; and (iii) $T \rightarrow P$ represents the triggering of convection and condensation. Consistent with ecohydrological theory, vegetation is assumed to influence rainfall indirectly through its impact on fluxes and thermodynamic state rather than via a direct pathway \cite{Bonan2019, Betts2023}. Based on this framework, our working hypothesis is that vegetation influences precipitation primarily through the pathways $V_{t-1} \rightarrow (F_{t-1}, T_{t-1}) \rightarrow P_t$. We therefore encode $T_{t-1}$ and $F_{t-1}$ as $A_{t-1}$ to represent atmospheric mediators.

To test whether the trained model relies on these pathways, we conducted a set of intervention-like input ablations at test time. Specifically, we retrained the model under three configurations: (1) removal of vegetation inputs ($V_{t-1}$), (2) removal of autoregressive precipitation ($P_{t-1}$), and (3) removal of both vegetation and precipitation memory. Performance differences across these configurations indicate whether vegetation contributes predictive information beyond precipitation persistence, and whether this contribution is expressed through atmospheric variables retained in the model.

Three configurations were tested: (1) \textit{do}($V_{t-1} = \emptyset$), removing all vegetation inputs; (2) \textit{do}($P_{t-1} = \emptyset$), removing autoregressive precipitation; and (3) \textit{do}($V_{t-1} = \emptyset, P_{t-1} = \emptyset$), removing both. If vegetation influences rainfall only via atmospheric mediators, then configuration (3) should degrade model performance more strongly than (2), providing evidence that $V \rightarrow F \rightarrow T \rightarrow P$ was learned. These ablations probe the internal dependency structure and mechanistic consistency of the trained neural network rather than providing formal causal identification of physical effects.

\subsubsection{Causal Graph Discovery}

To further evaluate whether the dependency structure learned by the model is consistent with the hypothesised indirect vegetation-atmosphere-precipitation pathway, we applied the PC (Peter-Clark) algorithm \cite{Kalisch2007}, a constraint-based causal discovery method that infers conditional independencies and outputs a Directed Graph (DG). Applied to temporally aligned, spatially aggregated data, the PC algorithm supported the hypothesised pathway $V \rightarrow F \rightarrow T \rightarrow P$, consistent with the physical process chain. No additional confounding variables were detected by this algorithm. Constraint-based causal discovery can only identify confounders that are present, varying and statistically resolvable within the dataset, which may not be detectable in our 1.5-year training period. This indicates that within the observed data and model representation, the learned dependence structure is consistent with an indirect vegetation-rainfall pathway mediated by atmospheric processes.

\subsubsection{Gradient Sensitivity Attribution}

To quantify the influence of vegetation on precipitation predictions, we compute the gradient of the loss $\mathcal{L}$ with respect to vegetation inputs $V_t$ at each timestep:

\[
g_t = \nabla_{V_t} \mathcal{L}(f(X_t), y_t), \quad g_t \in \mathbb{R}^{C \times H \times W}
\]

To identify the most influential linear combination of vegetation variables, we apply power iteration to extract the dominant direction $d^{(1)} \in \mathbb{R}^C$ that maximises the directional gradient norm:

\[
d^{(1)} = \arg\max_{\|d\| = 1} \left\| \sum_{i,j} g_t[:, i, j]^\top d \right\|^2
\]

This direction captures the vegetation variable combination with the greatest average gradient influence across space.

For further decomposition, we reshape $g_t$ into a matrix $G \in \mathbb{R}^{C \times (H \cdot W)}$ and compute its SVD:

\[
G = U \Sigma V^\top
\]

The left singular vectors $U$ represent orthogonal directions of vegetation sensitivity, and their singular values $\Sigma$ indicate relative importance. Temporal averaging of the top vectors across $t$ reveals patterns such as early dominance of evaporation and delayed influence of LAI and albedo. This gradient-based analysis identifies interpretable pathways of model sensitivity, supporting the presence of temporally staggered vegetation-precipitation couplings.

\subsection{Counterfactual Scenario Generation}

\subsubsection{Hotspot-Based Perturbation}

To produce a hypothetical of how deforestation spreads across the Amazon, we first identify hotspots, defined as areas where forest loss is most intense in recent years. Around these hotspots, we gradually perturb vegetation variables representing forest cover, so that the total vegetation loss matches realistic annual forest-decline rates. In this way, we can create counterfactual scenarios that ask: how would rainfall have changed if deforestation had been faster, slower, or spatially more concentrated?

Deforestation hotspots were extracted from the Hansen Global Forest Change v1.11 dataset via Google Earth Engine\footnote{\url{https://earthenginepartners.appspot.com/science-2013-global-forest}}. Pixels with 2023 forest loss (loss band = 23) were filtered using a minimum year-2000 tree cover threshold and clipped to the study region. The resulting raster was converted into a grid of hotspot centroids. To identify vegetation decline, we applied the non-parametric Mann-Kendall (MK) trend test to 2023 LAI time series, after first detrending with polynomial regression to remove global signals. The significance level was set at $\alpha = 0.05$. Grid cells with test statistics corresponding to $p < 0.05$ were classified as deforestation zones. A one-tailed test was used to assess negative trends only. Grid cells exhibiting significant negative trends were classified as deforestation zones. The MK test was chosen as it is distribution-free, robust to non-normality, and widely used for monotonic trend detection. 

A spatially decaying Gaussian mask was then applied:

\[
M_{ij} = \exp\left( -\frac{d_{ij}^2}{2\sigma^2} \right)
\]

Here, $d_{ij}$ is the Euclidean distance from grid cell $(i,j)$ to the nearest hotspot and $\sigma$ controls the spatial decay. The clipped mask ranged from 0.0 to 1.0.

Vegetation was perturbed as:

\[
\widetilde{V}_{ij}(t) = V_{ij}(t) + \rho \cdot t \cdot M_{ij}
\]

Where $\rho$ is the perturbation rate and $t$ is the number of elapsed hours. This approach simulates spatially localised deforestation radiating from roads or degraded areas.

To compute $\rho$, we solved:

\[
\delta_{\text{global}} \cdot L_0 = \rho \cdot t \cdot \left( f \cdot \overline{M}_{\text{hotspot}} + (1 - f) \cdot \overline{M}_{\text{non-hotspot}} \right)
\]

Where $\delta_{\text{global}}$ is the target LAI loss, $L_0$ the original mean LAI, and $f$ the hotspot area fraction.

Other vegetation variables (albedo, roughness, soil water, evaporation) were perturbed using their respective climatological statistics, scaled by the same $M_{ij}$ and $\rho$.

\begin{table}
\centering
\caption{Deforestation scenario parameters and estimated LAI change.}
\label{tab:deforestation_scenarios}
\begin{tabular}{p{4.5cm}p{2.5cm}p{2.5cm}p{2.5cm}}
\toprule
\textbf{Scenarios} & \textbf{Perturbation Rate} $\rho$ & \textbf{Decay Spread} $\sigma$ & \textbf{Estimated Forest Cover Change (\%)} \\
\midrule
\textit{Scenario 1: Heavy Diffuse Deforestation} & $-0.00014$ & $2.0$ & $-1.51$ \\
\textit{Scenario 2: Heavy Localied Deforestation}  & $-0.00014$ & $1.0$ & $-1.00$ \\
\textit{Scenario 3: Moderate Deforestation}       & $-0.000014$ & $2.0$ & $-0.38$ \\
\textit{Scenario 4: Low Deforestation}            & $-0.000010$ & $2.0$ & $-0.27$ \\
\textit{Scenario 5: Mild Regrowth}       & $+0.000004$ & $2.0$ & $+0.10$ \\
\bottomrule
\end{tabular}
\end{table}

\vspace{1em}

To explore a range of canopy change scenarios, we constructed five counterfactual deforestation experiments, seen in Table~\ref{tab:deforestation_scenarios}. These varied both the spatial spread ($\sigma$) and the perturbation rate ($\rho$), producing differing magnitudes and spatial patterns of LAI loss. Scenario names summarise the extent and intensity of vegetation change.

\vspace{1em}

\subsubsection{Analysis of Rainfall Prediction Changes and Hydrological Impact}

To evaluate rainfall prediction changes under each counterfactual vegetation scenario, we applied spatial and distributional analyses. Grid-wise difference maps and histograms visualised spatial and temporal shifts, while summary statistics (mean, minimum, maximum) quantified aggregate changes across the basin. To assess distributional changes beyond mean or variance shifts, we used the two-sample Kolmogorov-Smirnov (KS) test, stratified by rainfall intensity bands. This non-parametric test is well-suited to the heavy-tailed, non-normal nature of precipitation and measures the maximum gap between the cumulative distributions of baseline and perturbed predictions, highlighting structural differences in rainfall distributions.  

Formally, the KS statistic is defined as:  

\[
D = \sup_x |F_1(x) - F_2(x)|
\]

where \(F_1\) and \(F_2\) are the empirical cumulative distribution functions of the two samples. Statistical significance was evaluated using a two-tailed test at \(\alpha = 0.05\), with the critical threshold \(D_{\text{critical}}\) computed from the total sample size (\(n = 41{,}107{,}824\)). KS values exceeding this threshold were considered significant. The test was applied separately for each rainfall intensity bin to determine whether perturbations induced distributional shifts beyond sampling variability.   

In addition to the KS statistic, we applied three hydrological metrics to evaluate ecological and seasonal impacts. Relative Entropy (RE) was chosen to capture whether rainfall becomes more seasonally concentrated. Dry Season Intensity (DSI) was chosen to quantify ecological stress when evapotranspiration demand exceeds rainfall supply \cite{30, qin2025, flores2024, 27}. PAF was chosen to measure spatial heterogeneity, detecting whether the extent of the precipitating area expands or contracts under vegetation perturbations and studies suggest that PAF dominates the scaling of coarse-grained daily rainfall extremes \cite{roca2022}.

\begin{itemize}
    \item \textbf{Relative Entropy (RE)} quantifies how strongly rainfall is concentrated in particular months. Letting \( p_m = P_m / \sum_j P_j \) and \( q_m = 1/12 \) for monthly totals \( P_m \), we define \cite{pascale2014}:
    \[
    \text{RE} = \sum_{m=1}^{12} p_m \log_2\left(\frac{p_m}{q_m}\right). 
    \]

    \item \textbf{Dry Season Intensity (DSI)} measures cumulative daily water deficit, using daily rainfall \( P_d \) and potential evapotranspiration \( \text{PET}_d \):
    \[
    \text{DSI} = \sum_{d: P_d < \text{PET}_d} (\text{PET}_d - P_d),
    \]
    where PET was estimated using the Penman-Monteith equation. \cite{malhi2009, Allen2005ASCE, Penman1948, Monteith1965}

    \item \textbf{PAF} quantifies the spatial spread of rainfall:
    \[
    \text{PAF} = \frac{1}{H \cdot W} \sum_{i,j} \mathbb{1}_{\{y_{ij} > 0.1\}},
    \]
    representing the proportion of grid cells with rainfall above 0.1 mm/h. \cite{roca2022}
\end{itemize}

All metrics were computed monthly from hourly predictions to capture changes in seasonality, hydrological stress, and spatial distribution.

\subsubsection{Projected Gradient Descent for Sensitivity Attribution}

We use Projected Gradient Descent (PGD) to identify the minimum vegetation perturbations required to induce significant changes in predicted rainfall. PGD backpropagates prediction error with respect to vegetation inputs and iteratively updates them along the steepest gradient direction, subject to ecologically realistic bounds. This yields spatially explicit maps of rainfall sensitivity and identifies the vegetation variables to which precipitation is most responsive.

PGD was applied to six vegetation-related variables: high/low LAI, surface roughness, soil water content, albedo, and evaporation. 

PGD was conducted both with and without spatial constraints to historical deforestation hotspots. Hotspots were derived from the Hansen Global Forest Change dataset and Mann-Kendall trend analysis on annual LAI, as described in Section 4.4.1. In the hotspot case, a Gaussian decay mask was applied to weight perturbations spatially:

\[
M(x, y) = \exp \left( -\frac{d^2(x, y)}{2\sigma^2} \right)
\]

where \( d(x, y) \) is the Euclidean distance to the nearest hotspot and \( \sigma = 2 \). The mask was clipped to the range \([0.0, 1.0]\) to maintain locality.

At each PGD step, vegetation variables were updated as:

\[
\delta X = \text{proj}_{[\Delta v_{\min}, \Delta v_{\max}]} \left( \alpha \cdot M \cdot d' \right)
\]

Here, \( d' \) is the dominant gradient direction (estimated via power iteration; see Section~4.3.3), \( \alpha = 0.005 \) is the learning rate, and \( \delta X \) is the perturbation vector. The dominant direction \( d' \) was computed from the loss gradient, capturing the steepest ascent in the vegetation subspace. This vector was averaged over 30-day windows to estimate monthly sensitivities. The projection operator enforces physically realistic limits on hourly change:

\[
\Delta v_{\max} = \frac{v_{\max} - v_{\min}}{365 \cdot 24}
\]

These per-variable, per-grid bounds were computed from climatological minima and maxima, ensuring ecologically plausible perturbations.

We tracked the mean absolute contribution of each variable to precipitation changes to understand which vegetation features most consistently influenced rainfall. This produced sensitivity maps of precipitation to each input variable.

The pretrained MultiTask ConvLSTM model predicted rainfall at each iteration, and PGD steps continued until the model exhibited a statistically significant change: either a mean rainfall shift exceeding 0.1 mm/h or a change in PAF above 5\%. In convective systems like the Amazon, rainfall may shift spatially without major intensity change; hence, PAF, defined as the fraction of grid cells with rainfall $>$ 0.1 mm/h, captures changes in spatial extent.

The full PGD objective was formulated as:

\[
\min_{\delta X} \mathcal{L}(f(X + \delta X), y) \quad \text{subject to} \quad \delta X = \text{proj}_{[\Delta v_{\min}, \Delta v_{\max}]} \left( \alpha \cdot M \cdot d' \right)
\]

where \( \mathcal{L} \) is the loss function (MSE), \( f \) the pretrained ConvLSTM model, \( X \) the original vegetation input, and \( y \) the unperturbed rainfall target. This setup facilitates the quantification of the minimum vegetation perturbation required to induce a significant rainfall change.

\section*{Data Availability}

All data used in this study are publicly available. ERA5-Land hourly reanalysis data on single levels from 1940 to present were obtained from the Copernicus Climate Data Store (\url{https://cds.climate.copernicus.eu/cdsapp#!/dataset/reanalysis-era5-single-levels}). Forest cover loss data were retrieved from the Hansen Global Forest Change v1.11 dataset via Google Earth Engine (\url{https://earthenginepartners.appspot.com/science-2013-global-forest}).

\section*{Code Availability}

The code supporting this study is available from the corresponding author upon reasonable request. The trained MultiTask ConvLSTM is available on Hugging Face, with metrics reported on the validation set: \url{https://huggingface.co/makkos-lilly/MultiTaskConvLSTM}.

\section*{Acknowledgements}
The author received no external funding for this work. 

\section*{Author Contributions}
L.H.-M. designed and conducted the research and wrote the manuscript. F.M. supervised the project, provided feedback on the research design and interpretation, and reviewed the manuscript.

\section*{Competing Interests}
The authors declare no competing interests.

\section{Additional Information}
Correspondence and requests for materials should be addressed to L.H.-M.

\FloatBarrier

\newpage

\bibliographystyle{naturemag}  
\bibliography{references}

@article{bagley2014,
  author    = {Bagley, J. and Desai, A. and Harding, K. and Synder, P. and Foley, J.},
  title     = {Drought and deforestation: Has land cover change influenced recent precipitation extremes in the Amazon?},
  journal   = {Journal of Climate},
  volume    = {27},
  number    = {1},
  pages     = {345-361},
  year      = {2014},
  doi       = {10.1175/jcli-d-12-00369.1}
}

@article{badger2015,
  author    = {Badger, A. M. and Dirmeyer, P. A.},
  title     = {Remote tropical and sub-tropical responses to Amazon deforestation.},
  journal   = {Climate Dynamics},
  volume    = {46},
  number    = {9-10},
  pages     = {3057--3066},
  year      = {2015},
  doi       = {10.1007/s00382-015-2752-5}
}

@article{butt2023,
  author    = {Butt, E. and Baker, J. and Gilney, F. and von Randow, C. and Paula, A. and Spracklen, D. V.},
  title     = {Amazon deforestation causes strong regional warming.},
  journal   = {Proceedings of the National Academy of Sciences of the United States of America},
  volume    = {120},
  number    = {45},
  pages     = {},
  year      = {2023},
  doi       = {10.1073/pnas.2309123120}
}

@article{konapala2020,
  author    = {Konapala, G. and Mishra, A. K. and Wada, Y. and Mann, M. E.},
  title     = {Climate change will affect global water availability through compounding changes in seasonal precipitation and evaporation.},
  journal   = {Nature Communications},
  volume    = {11},
  number    = {1},
  pages     = {3044},
  year      = {2020},
  doi       = {10.1038/s41467-020-16757-w},
}

@article{lawrence2015,
  author    = {Lawrence, D. and Vandecar, K.},
  title     = {Effects of tropical deforestation on climate and agriculture.},
  journal   = {Nature Climate Change},
  volume    = {5},
  number    = {1},
  pages     = {27--36},
  year      = {2015},
  doi       = {10.1038/nclimate2430},
}

@article{mu2022,
  author    = {Mu, Y. and Jones, C.},
  title     = {An observational analysis of precipitation and deforestation age in the Brazilian Legal Amazon.},
  journal   = {Atmospheric Research},
  volume    = {271},
  number    = {},
  pages     = {106122},
  year      = {2022},
  doi       = {10.1016/j.atmosres.2022.106122}
}

@article{leitefilho2021,
  author    = {Leite-Filho, A. T. and Soares-Filho, B. S. and Davis, J. L. and Abrahão, G. M. and Börner, J.},
  title     = {Deforestation reduces rainfall and agricultural revenues in the Brazilian Amazon.},
  journal   = {Nature Communications},
  volume    = {12},
  number    = {1},
  pages     = {},
  year      = {2021},
  doi       = {10.1038/s41467-021-22840-7}
}

@article{duku2023,
  author    = {Duku, C. and Hein, L.},
  title     = {Assessing the impacts of past and ongoing deforestation on rainfall patterns in South America.},
  journal   = {Global Change Biology},
  volume    = {29},
  number    = {18},
  pages     = {},
  year      = {2023},
  doi       = {10.1111/gcb.16856}
}

@article{qin2025,
  author    = {Qin, Y. and Wang, D. and Ziegler, A. D. and Fu, B. and Zeng, Z.},
  title     = {Impact of Amazonian deforestation on precipitation reverses between seasons.},
  journal   = {Nature},
  volume    = {639},
  number    = {8053},
  pages     = {102--108},
  year      = {2025},
  doi       = {10.1038/s41586-024-08570-y}
}

@article{dacruz2021,
  author    = {da Cruz, D. C. and Benayas, J. M. R. and Ferreira, G. C. and Santos, S. R. and Schwartz, G.},
  title     = {An overview of forest loss and restoration in the Brazilian Amazon.},
  journal   = {New Forests},
  volume    = {52},
  number    = {52},
  pages     = {1--16},
  year      = {2020},
  doi       = {10.1007/s11056-020-09777-3}
}

@article{14,
  author    = {Miller, A. J.},
  title     = {Remote sensing proxies for deforestation and soil degradation in landslide mapping: A review.},
  journal   = {Geography Compass},
  volume    = {7},
  number    = {7},
  pages     = {489--503},
  year      = {2013},
  doi       = {10.1111/gec3.12050}
}

@article{15,
  author    = {Hou, Y. and Wei, X. and Zhang, M. and Creed, I. F. and McNulty, S. G. and Ferraz, S. F. B.},
  title     = {A global synthesis of hydrological sensitivities to deforestation and forestation.},
  journal   = {Forest Ecology and Management},
  volume    = {529},
  issue     = {},
  pages     = {120718},
  year      = {2023},
  doi       = {10.1016/j.foreco.2022.120718}
}

@article{16,
  author    = {Dirmeyer, P. A. and Shukla, J.},
  title     = {Albedo as a modulator of climate response to tropical deforestation.},
  journal   = {Journal of Geophysical Research},
  volume    = {99},
  number    = {D10},
  pages     = {20863},
  year      = {1994},
  doi       = {10.1029/94JD01311}
}

@article{17,
  author    = {Giambelluca, T. W. and Holscher, D. and Bastos, T. X. and Frazao, R. R. and Nullet, M. A. and Ziegler, A. D.},
  title     = {Observations of albedo and radiation balance over postforest land surfaces in the eastern Amazon Basin.},
  journal   = {Journal of Climate},
  year      = {1997},
  volume    = {10},
  number    = {5},
  pages     = {919--928},
  doi       = {10.1175/1520-0442(1997)010<0919:ooaarb>2.0.co;2}
}

@article{18,
  author    = {Berbet, M. L. C. and Costa, M. H.},
  title     = {Climate change after tropical deforestation: Seasonal variability of surface albedo and its effects on precipitation change.},
  journal   = {Journal of Climate},
  volume    = {16},
  number    = {12},
  pages     = {2099--2104},
  year      = {2003},
  doi       = {10.1175/1520-0442(2003)016<2099:CCATDS>2.0.CO;2}
}

@article{19,
  author    = {Mgelwa, A. S. and Ngaba, M. J. Y. and Hu, B. and Gurmesa, G. A. and Mwakaje, A. G. and Nyemeck, M. P. B. and Zhu, F. and Qiu, Q. and Song, L. and Wang, Y. and Fang, Y. and Rennenberg, H.},
  title     = {Meta-analysis of 21st century studies shows that deforestation induces profound changes in soil characteristics, particularly soil organic carbon accumulation.},
  journal   = {Forest Ecosystems},
  year      = {2024},
  volume    = {12},
  number    = {},
  pages     = {100257},
  doi       = {10.1016/j.fecs.2024.100257}
}

@article{20,
  author    = {Winckler, J. and Reick, C. H. and Bright, R. M. and Pongratz, J.},
  title     = {Importance of surface roughness for the local biogeophysical effects of deforestation.},
  journal   = {Journal of Geophysical Research: Atmospheres},
  volume    = {124},
  number    = {15},
  pages     = {8605--8618},
  year      = {2019},
  doi       = {10.1029/2018JD030127}
}

@inproceedings{21,
  author = {Shi, X. and Chen, Z. and Wang, H. and Yeung, D.-Y. and Wong, W.-K and & Woo, W.-C.},
  title = {Convolutional LSTM network: A machine learning approach for precipitation nowcasting},
  booktitle = {Adv. Neural Inf. Process. Syst.},
  pages = {802--810},
  year = {2015},
  url = {https://arxiv.org/abs/1506.04214}
}

@article{27,
  author    = {Zhang, S. and Yang, Y. and McVicar, T. R. and Yang, D.},
  title     = {An analytical solution for the impact of vegetation changes on hydrological partitioning within the Budyko framework.},
  journal   = {Water Resources Research},
  volume    = {54},
  number    = {1},
  pages     = {519--537},
  year      = {2018},
  doi       = {10.1002/2017WR022028}
}

@article{29,
  author    = {Zemp, D. C. and Schleussner, C.-F. and Barbosa, H. M. J. and Ramming, A.},
  title     = {Deforestation effects on Amazon forest resilience.},
  journal   = {Geophysical Research Letters},
  volume    = {44},
  number    = {12},
  pages     = {6182--6190},
  year      = {2017},
  doi       = {10.1002/2017GL072955}
}

@article{30,
  author    = {Medvigy, D. and Walko, R. L. and Avissar, R.},
  title     = {Effects of Deforestation on Spatiotemporal Distributions of Precipitation in South America.},
  journal   = {Journal of Climate},
  volume    = {24},
  number    = {8},
  pages     = {2147--2163},
  year      = {2011},
  doi       = {10.1175/2010JCLI3882.1}
}

@article{extreme_mean_precip,
  author    = {Zhang, M. and Gao, Y. and Ge, J.},
  title     = {Different responses of extreme and mean precipitation to land use and land cover changes},
  journal   = {npj Climate and Atmospheric Science},
  volume    = {8},
  number    = {175},
  pages     = {},
  year      = {2025},
  doi       = {10.1038/s41612-025-01049-1}
}

@article{4dvar,
  author    = {Wang, W. and Zhang, J. and Su, Q. and Chai, X. and Lu, J. and Ni, W. and Duan, B. and Ren, K.},
  title     = {Accurate initial field estimation for weather forecasting with a variational constrained neural network.},
  journal   = {npj Climate and Atmospheric Science},
  year      = {2024},
  volume    = {7},
  number    = {1},
  pages     = {},
  doi       = {10.1038/s41612-024-00776-1}
}

@article{transformer,
  author    = {Lu, C. and Shen, Y. and Guan. Z.},
  title     = {A modified transformer model for the extended-range forecast of intraseasonal oscillation.},
  journal   = {npj Climate and Atmospheric Science},
  year      = {2025},
  volume    = {8},
  number    = {1},
  pages     = {},
  doi       = {10.1038/s41612-025-00902-7}
}

@article{convlstmhybrid,
  author    = {Moishin, M. and Deo, R. C. and Prasad, R. and Raj, N. and Abdulla, S.},
  title     = {Designing deep-based learning flood forecast model with convlstm hybrid algorithm.},
  journal   = {IEEE Access},
  volume    = {9},
  number    = {},
  pages     = {50982--50993},
  year      = {2021},
  doi       = {10.1109/access.2021.3065939}
}

@article{CNN_convective,
  author    = {Caseri, A. and Santos, L. L. and Stephany, S.},
  title     = {A convolutional recurrent neural network for strong convective rainfall nowcasting using weather radar data in Southeastern Brazil.},
  journal   = {Artificial Intelligence in Geosciences},
  volume    = {},
  number    = {},
  pages   = {},
  year      = {2022},
  doi       = {10.1016/j.aiig.2022.06.001.}
}

@article{liao2024short,
  author = {L, Y. and Lu, S. and Yin, G.},
  title = {Short-term and imminent rainfall prediction model based on ConvLSTM and SMAAT-UNet.},
  journal = {Sensors},
  volume = {24},
  number = {11},
  pages = {3576},
  year = {2024},
  doi = {10.3390/s24113576}
}

@article{wang2021multi,
  author = {Du, X. and Guo, H},
  title = {A multi-scale attention encoding and dynamic decoding network designed for short-term precipitation forecasting.},
  journal = {Earth Science Informatics},
  volume  = {18},
  number  = {1},
  pages = {},
  year = {2024},
  doi = {10.1007/s12145-024-01554-6}
}

@article{xu2024,
  author = {Xu L. and Zhang, X. and Yu, H. and Chen, Z. and Du, W. and Chen, N.},
  title = {Incorporating spatial autocorrelation into deformable ConvLSTM for hourly precipitation forecasting.},
  journal = {Computers \& Geosciences},
  volume = {184},
  number = {},
  pages = {105536},
  year = {2024},
  doi = {10.1016/j.cageo.2024.105536}
}

@article{rui2024,
  author = {Rui, C. and Sun, Z. and Zhang, W. and Liu, A. and Wei, Z.},
  title = {Enhancing ENSO predictions with self-attention ConvLSTM and temporal embeddings.},
  journal = {Frontiers in Marine Science},
  volume = {11},
  number = {},
  pages = {},
  year = {2024},
  doi = {10.3389/fmars.2024.1334210}
}

@article{smith2023,
  author = {C. Smith and J. C. A. Baker and D. V. Spracklen},
  title = {Tropical deforestation causes large reductions in observed precipitation.},
  journal = {Nature},
  volume = {615},
  number = {},
  pages = {270--275},
  year = {2023},
  doi = {10.1038/s41586-022-05690-1}
}

@article{medvigy2011,
  author = {Avissar, R. and Walko, R. and Medvigy, D.},
  title = {Effects of deforestation on spatiotemporal distributions of precipitation in south america.},
  journal = {Journal of Climate},
  volume = {P24},
  number = {8},
  pages = {2147--2163},
  year = {2011},
  doi = {0.1175/2010JCLI3882.1}
}

@article{Lapuschkin2019,
  title={Unmasking Clever Hans predictors and assessing what machines really learn.},
  author={Lapuschkin, S. and Waldchen, S. and Binder, A. and Montavon, G. and Samek, W. and Muller, K-R.},
  journal={Nature Communications},
  volume={10},
  number={1},
  pages={1096},
  year={2019},
  doi = {10.1038/s41467-019-08987-4},
}

@article{Zhou2022,
  title={Do feature attribution methods correctly attribute features?},
  author={Zhou, Y. and Booth, S. and Ribeiro, M. T. and Shah, J.},
  journal={Proceedings of the AAAI Conference on Artificial Intelligence},
  volume={36},
  number={9},
  pages={9623--9633},
  year={2022},
  doi={10.1609/aaai.v36i9.21196}
}

@article{Runge2019,
  title={Inferring causation from time series in Earth system sciences.},
  author={Runge, J. and Bathiany, S. and Bollt, E. and Camps-Valls, G. and Coumou, D. and Deyle, E. and Glymour, C. and Kretschmer, M. and Mahecha, M. D. and Munoz-Mari, J. and van Nes, E. H. and Peters, J. and Quax, R. and Reichstein, M. and Scheffer, M. and Scholkopf, B. and Spirtes, P. and Sugihara, G. and Sun, J. and Zhang, K. and Zscheischler, J.},
  journal={Nature Communications},
  volume={10},
  number={1},
  pages={2553},
  year={2019},
  doi={10.1038/s41467-019-10105-3}
}

@book{Pearl2009,
  title={Causality: Models, Reasoning and Inference.},
  author={Pearl, Judea},
  year={2009},
  publisher={Cambridge University Press},
  address = {Cambridge},
}

@article{Ravuri2021,
  title={Skilful precipitation nowcasting using deep generative models of radar.},
  author={Ravuri, S. and Lenc, K. and Willson, M. and Kangin, D. and Lam, R. and Mirowski, P. and Fitzsimons, M. and Athanassiadou, M. and Kashem, S. and Madge, S. and Prudden, R. and Mandhane, A. and Clark, A. and Brock, A. and Simonyan, K. and Hadsell, R. and Robinson, N. and Clancy, E. and Arribas, A. and Mohamed, S.},
  journal={Nature},
  volume={597},
  number={7878},
  pages={672--677},
  year={2021},
  doi ={10.1038/s41586-021-03854-z}
}

@misc{Sonderby2020,
      title={MetNet: A Neural Weather Model for Precipitation Forecasting.}, 
      author={Sønderby, C. K. and Espeholt, L. and Heek, J. and Dehghani, M. and Oliver, A. and Salimans, T. and Agrawal, S. and Hickey, J. and Kalchbrenner, N.},
      year={2020},
      eprint={2003.12140},
      archivePrefix={arXiv},
      primaryClass={cs.LG},
      url={https://arxiv.org/abs/2003.12140}, 
}

@article{espeholt2022,
  title={Deep learning for twelve hour precipitation forecasts.},
  author={Espeholt, L. and Agrawal, S. and Sonderby, C. and Kumar, M. and Heek, J. and Bromberg, C. and Gazen, C. and Carver, R. and Andrychowicz, M. and Hickey, J. and Bell, A. and Kalchbrenner, N.},
  journal={Nature Communications},
  volume={13},
  number={1},
  pages={5145},
  year={2022},
  doi ={10.1038/s41467-022-32483-x}
}

@article{flores2024,
  title={Critical transitions in the Amazon forest system.},
  author={Flores, B. M. and Montoya, E. and Sakschewski, B. and Nascimento, N. and Staal, A. and Betts, R. A. and Levis, C. and Lapola, D. M. and Esquível-Muelbert, A. and Jakovac, C. and Nobre, C. A. and Oliveira, R. S. and Borma, L. S. and Nian, D. and Boers, N. and Hecht, S. B. and ter Steege, H. and Arieira, J. and Lucas, I. L. and Berenguer, E. and Marengo, J. A. and Gatti, L. V. and Mattos, C. R. C. and Hirota, M.},
  journal={Nature},
  volume={626},
  number={7999},
  pages={555–-564},
  year={2024},
  doi ={10.1038/s41586-023-06970-0}
}

@article{doughty2012,
  title= {Theoretical Impact of Changing Albedo on Precipitation at the Southernmost Boundary of the ITCZ in South America.},
  author={Doughty, C.E. and Loarie, S.R. and Field, C.B. },
  journal={Earth Interactions},
  volume={16},
  number={8},
  pages={1–-14},
  year={2012},
  doi ={10.1175/ei422.1}
}

@article{sanchezmartinez2025,
  title={Amazon rainforest adjusts to long-term experimental drought.},
  author={Sanchez-Martinez, P. and Martius, L. R. and Bittencourt, P. and Silva, M. and Binks, O. and Coughlin, I. and Negrao-Rodrigues, V. and Silver, J. A. and Da Costa, A. C. L. and Selman, R. and Rifai, S. and Rowland, L. and Mencuccini, M. and Meir, P.},
  journal={Nature Ecology \& Evolution},
  volume = {},
  number = {},
  pages = {},
  year={2025},
  doi ={10.1038/s41559-025-02702-x}
}

@article{lovo2024,
  title={Tackling the Accuracy-Interpretability Trade-off in a Hierarchy of Machine Learning Models for the Prediction of Extreme Heatwaves.},
  author={Lovo, A. and Lancelin, A. and Herbert, C. and Bouchet, F.},
  journal = {Artificial Intelligence Earth Systems},
  volume = {4},
  number = {},
  pages = {240094},
  year={2025},
  doi={10.1175/AIES-D-24-0094.1}
}

@misc{dziugaite2020,
      title={Enforcing Interpretability and its Statistical Impacts: Trade-offs between Accuracy and Interpretability}, 
      author={Gintare K. D. and Ben-David, S. and Roy, D. M},
      year={2020},
      eprint={2010.13764},
      archivePrefix={arXiv},
      primaryClass={cs.LG},
      url={https://arxiv.org/abs/2010.13764}, 
}

@article{2zemp2017,
  title={Self-amplified Amazon forest loss due to vegetation-atmosphere feedbacks.},
  author={Zemp, D. C. and Schleussner, C-F. and Barbosa, H. MJ. and Hirota, M. and Montade, V. and Sampaio, G. and Staal, A. and Wang-Erlandsson, L. and Rammig, A.},
  journal={Nature Communications},
  volume={8},
  number={},
  pages={14681},
  year={2017},
  doi={10.1038/ncomms14681}
}

@article{boers2017,
  title={A deforestation-induced tipping point for the South American monsoon system.},
  author={Boers, N. and Marwan, N. and Barbosa, H. M. J. and Kurths, J.},
  journal={Scientific Reports},
  volume={7},
  number={},
  pages={41489},
  year={2017},
  doi={10.1038/srep41489}
}

@inproceedings{paszke2019pytorch,
  title={PyTorch: An Imperative Style, High-Performance Deep Learning Library.},
  author={Paszke, A. and Gross, S. and Massa, F. and Lerer, A. and Bradbury, J. and Chanan, G. and Killeen, T. and Lin, Z. and Gimelshein, N. and Antiga, L. and Desmaison, A. and Köpf, A. and Yang, E. and DeVito, Z. and Raison, M. and Tejani, A. and Chilamkurthy, S. and Steiner, B. and Fang, L. and Bai, J. and Chintala, S.},
  booktitle={Advances in Neural Information Processing Systems 32},
  pages={8024--8035},
  year={2019}
}

@article{fang2019,
  title={An overview of global leaf area index (LAI): methods, products, validation, and applications.},
  author={Fang, H. and Baret, F. and Plummer, S. and Schaepman-Strub, G.},
  journal={Reviews of Geophysics},
  volume={57},
  number={3},
  pages={739-799},
  year={2019},
  doi={10.1029/2018RG000608}
}

@article{ma2019,
  title={An Advanced Multiple-Layer Canopy Model in the WRF Model With Large-Eddy Simulations to Simulate Canopy Flows and Scalar Transport Under Different Stability Conditions.},
  author={Ma, Y. and Liu, H.},
  journal={Journal of Advances in Modeling Earth Systems},
  volume={11},
  number={7},
  pages={2330-2351},
  year={2019},
  doi={10.1029/2018ms001347}
}

@article{zhang2022,
  title={Review of Land Surface Albedo: Variance Characteristics, Climate Effect and Management Strategy.},
  author={Zhang, X. AND Jiao, Z. and Zhao, C. and Qu, Y. and Lui, Q. and Zhang, H. and Tong, Y. and Wang, C. and Li, S. and Guo, J. and Zhu, Z. and Yin, S. and Cui, L.},
  journal={Remote Sensing},
  volume={14},
  number={6},
  pages={1382},
  year={2019},
  doi={10.3390/rs14061382}
}

@article{lejeune2014,
  title={Influence of Amazonian deforestation on the future evolution of regional surface fluxes, circmulations, surface temperature and precipitation.},
  author={Lejeune, Q. and Davin, E. L. and Guillod, B. P. and Seneviratne, S. I.},
  journal={Climate Dynamics},
  volume={44},
  number={9-10},
  pages={2769-2786},
  year={2014},
  doi={10.1007/s00382-014-2203-8}
}

@article{li2025,
  title={Extreme Precipitation Nowcasting Using Multitask Latent Diffusion Models.},
  author={Li, C. and Ling, X. and Yang, Q. and Chen, M. and Qin, F. and Huang, Y.},
  journal={IEEE Transactions on Geoscience and Remote Sensing},
  volume={63},
  number={},
  pages={1-15},
  year={2025},
  doi={10.48550/arXiv.2410.14103}
}

@article{nian2024,
  title={Rainfall seasonality domainates critical precipitation threshold for the Amazon forest in the LPJmL vegetation model.},
  author={Nian, D. and Bathiany, S. and Sakschewski, B. and Druke, M. and Blaschke, L. and Ben-Yami, M. and von Bloh, W. and Boers, N.},
  journal={Science of The Total Environment},
  volume={947},
  number={},
  pages={174378},
  year={2024},
  doi={10.1016/j.scitotenv.2024.174378}
}

@article{chavez2017,
  title={Orographic rainfall hot spots in the Andes-Amazon transition according to the TRMM precipitation radar and in situ data.},
  author={Chavez, S. P. and Takahashi, K.},
  journal={Journal of Geophysical Research: Atmospheres},
  volume={122},
  number={11},
  pages={5870-5882},
  year={2024},
  doi={10.1002/2016JD026282}
}

@article{staal2020,
  title={Hysteresis of tropical forests in the 21st century.},
  author={Staal, A and Fetzer, I. and Wang-Erlandsson, L. and Bosmans, J. H. C and Dekker, S. C. and van Nes, E. H. and Rockstrom, J. and Tuinenburg, O. A.},
  journal={Nature Communications},
  volume={11},
  number={1},
  pages={},
  year={2020},
  doi={10.1038/s41467-020-18728-7}
}

@article{sham2025,
  title={
Advances in AI-based rainfall forecasting: a comprehensive review of past, present, and future directions with intelligent data fusion and climate change models.},
  author={Sham, F. A. F. and El-Shafie, A. and Wan Jaafar, W. Z. and Sankaran, A. and Sherif, M. and Ahmed, A. M. A. N.},
  journal={Results in Engineering},
  volume={27},
  number={},
  pages={105774},
  year={2025},
  doi={10.1016/j.rineng.2025.105774}
}

@article{roca2022,
  title={Precipitating Fraction, Not Intensity, Explains Extreme Coarse-Grained Precipitation Clausius-Clapeyron Scaling With Sea Surface Temperature Over Tropical Oceans.},
  author={Roca, R. and De Meyer, V. and Muller, C.},
  journal={Geophysical Research Letters},
  volume={49},
  number={24},
  pages={},
  year={2020},
  doi={10.1029/2022gl100624}
}

@article{pascale2014,
  title={Analysis of rainfall seasonality from observations and climate models.},
  author={Pascale, S. and Lucarini, V. and Feng, X. and Porporato, A. and ul Hasson, S.},
  journal={Climate Dynamics},
  volume={44},
  number={11-12},
  pages={},
  year={2014},
  doi={10.1007/s00382-014-2278-2}
}

@article{malhi2009,
  title={Exploring the likelihood and mechanism of climate-change-induced dieback of the Amazon rainforest.},
  author={Malhi, Y and Aragao, L. E. O. C. and Galbraith, D. and Huntingford, C. and Fisher, R. and Zelazowski, P. and Sitch, S. and McSweeney, C. and Meir, P.},
  journal={Proceedings of the National Academy of Sciences},
  volume={106},
  number={49},
  pages={20610-20615},
  year={2009},
  doi={10.1073/pnas.0804619106}
}

@article{Allen2005ASCE,
  author  = {Allen, Richard G. and Walter, I. A. and Elliott, R. L. and Howell, T. A. and Itenfisu, D. and Jensen, M. E.},
  title   = {The ASCE Standardized Reference Evapotranspiration Equation},
  journal = {Journal of Irrigation and Drainage Engineering},
  year    = {2005},
  volume  = {131},
  number  = {1},
  pages   = {63--72},
  doi     = {10.1061/(ASCE)0733-9437(2005)131:1(63)}
}

@article{Penman1948,
  author  = {Penman, H. L.},
  title   = {Natural Evaporation from Open Water, Bare Soil and Grass},
  journal = {Quarterly Journal of the Royal Meteorological Society},
  year    = {1948},
  volume  = {74},
  number  = {294},
  pages   = {24--46},
  doi     = {10.1002/qj.49707429403}
}

@incollection{Monteith1965,
  author    = {Monteith, J. L.},
  title     = {Evaporation and Environment},
  booktitle = {The State and Movement of Water in Living Organisms},
  series    = {Symposia of the Society for Experimental Biology},
  volume    = {19},
  pages     = {205--234},
  year      = {1965},
  publisher = {Cambridge University Press}
}

@incollection{Bonan2019,
  author    = {Bonan, G. B.},
  title     = {Turbulent Fluxes and Scalar Profiles in the Surface Layer},
  booktitle = {Climate change and terrestrial ecosystem modeling},
  series    = {Symposia of the Society for Experimental Biology},
  volume    = {1},
  pages     = {80--100},
  year      = {2019},
  publisher = {Cambridge University Press}
}

@article{Betts2023,
  author  = {Betts, A. K.},
  title   = {Understanding Hydrometeorology Using Global Models},
  journal = {Bulletin of the American Meteorological Society},
  year    = {2004},
  volume  = {85},
  number  = {111},
  pages   = {1673-1688},
  doi     = {10.1175/bams-85-11-1673}
}

@article{Kalisch2007,
  author  = {Kalisch, M. and Buhlmann, P.},
  title   = {Estimating High-Dimensional Directed Acyclic Graphs with the PC-Algorithm},
  journal = {Journal of Machine Learning Research},
  year    = {2007},
  volume  = {},
  number  = {},
  pages   = {},
  doi     = {10.5555/1248659.1248681}
}

@article{baudenaEffectsLanduseChange2021,
  title = {Effects of Land-Use Change in the {{Amazon}} on Precipitation Are Likely Underestimated},
  author = {Baudena, Mara and Tuinenburg, Obbe A. and Ferdinand, Pendula A. and Staal, Arie},
  date = {2021},
  journaltitle = {Global Change Biology},
  volume = {27},
  number = {21},
  pages = {5580--5587},
  issn = {1365-2486},
  doi = {10.1111/gcb.15810},
  url = {https://onlinelibrary.wiley.com/doi/abs/10.1111/gcb.15810},
  urldate = {2026-04-17}
}

@article{dukuAssessingImpactsOngoing2023,
  title = {Assessing the Impacts of Past and Ongoing Deforestation on Rainfall Patterns in {{South America}}},
  author = {Duku, Confidence and Hein, Lars},
  date = {2023},
  journaltitle = {Global Change Biology},
  volume = {29},
  number = {18},
  pages = {5292--5303},
  issn = {1365-2486},
  doi = {10.1111/gcb.16856},
  url = {https://onlinelibrary.wiley.com/doi/abs/10.1111/gcb.16856},
  urldate = {2026-04-17}
}

@article{esquivel-muelbertCompositionalResponseAmazon2019,
  title = {Compositional Response of {{Amazon}} Forests to Climate Change},
  author = {Esquivel-Muelbert, Adriane and Baker, Timothy R. and Dexter, Kyle G. and Lewis, Simon L. et al.},
  date = {2019},
  journaltitle = {Global Change Biology},
  volume = {25},
  number = {1},
  pages = {39--56},
  issn = {1365-2486},
  doi = {10.1111/gcb.14413},
  url = {https://onlinelibrary.wiley.com/doi/abs/10.1111/gcb.14413},
  urldate = {2026-04-17}
}

\clearpage

\clearpage 

\section*{Supplementary Note 1: ConvLSTM Equations}
We present the ConvLSTM update equations used in our framework. \vspace{0.5mm}
Each ConvLSTM cell updates its hidden state $H_t$ and memory cell $C_t$ as follows:
\begin{align*}
i_t &= \sigma(W_{xi} \ast X_t + W_{hi} \ast H_{t-1} + b_i) \\
f_t &= \sigma(W_{xf} \ast X_t + W_{hf} \ast H_{t-1} + b_f) \\
C_t &= f_t \odot C_{t-1} + i_t \odot \tanh(W_{xc} \ast X_t + W_{hc} \ast H_{t-1} + b_c) \\
o_t &= \sigma(W_{xo} \ast X_t + W_{ho} \ast H_{t-1} + b_o) \\
H_t &= o_t \odot \tanh(C_t)
\end{align*}
where $\ast$ is convolution, $\odot$ is the Hadamard product, and $\sigma$ is the sigmoid function. \cite{21}

\section*{Supplementary Note 2: Hyperparameter Selection and Training Protocol}

Hyperparameters were tuned via 5-fold cross-validation. Candidate loss functions included standard MSE, Balanced MSE, and Smooth L1; Balanced MSE yielded the best performance for low-intensity rainfall and was selected.

For precipitation, we evaluated z-score, min-max, log transforms (offsets $1$ and $10^{-6}$), and log-standardisation. Log normalisation with offset $1$ performed best. All other variables were standardised using z-score normalisation.

Architectural tuning included hidden dimensions for the last two ConvLSTM layers {(32, 64), (32, 128), (64, 128)}, with (32, 64) chosen for optimal capacity-efficiency balance. Kernel sizes $(2 \times 2)$, $(3 \times 3)$, and $(5 \times 5)$ were tested; $(3 \times 3)$ was selected for spatial resolution and parameter efficiency.

Final choices were based on average validation performance and metric variance, with preference given to stable configurations.

The model consists of a shared 3-layer ConvLSTM encoder, followed by a 3D convolutional layer that refines the learned spatiotemporal features before branching into two task-specific heads. The classification head is a Conv3D layer that outputs the probability of dry conditions, while the regression head is a 3-layer MLP that estimates the intensity of rain.

The input at each timestep is a tensor of atmospheric and vegetation variables from the most recent hour, with shape $[(T=1), H, W, C]$. These inputs are processed through three ConvLSTM layers with kernel size $(3 \times 3)$ and hidden dimensions of 14, 32, and 64. Each layer is followed by batch normalisation to improve training stability. A dropout layer is applied after the final MLP to reduce overfitting. Both output heads receive gradient updates during training, allowing the encoder to learn shared spatiotemporal representations relevant to both rainfall occurrence and intensity.

The model is trained using the composite loss described in Section 2.2.1, combining classification and regression terms. Training was conducted for 200 epochs and completed in approximately 45 minutes on a single GPU.

Optimisation is performed using the AdamW optimiser with an initial learning rate of 0.005, regulated by PyTorch's \textit{ReduceLROnPlateau} learning rate scheduler \cite{paszke2019pytorch}.

\begin{figure}[t]
    \centering

    \includegraphics[width=1\textwidth]{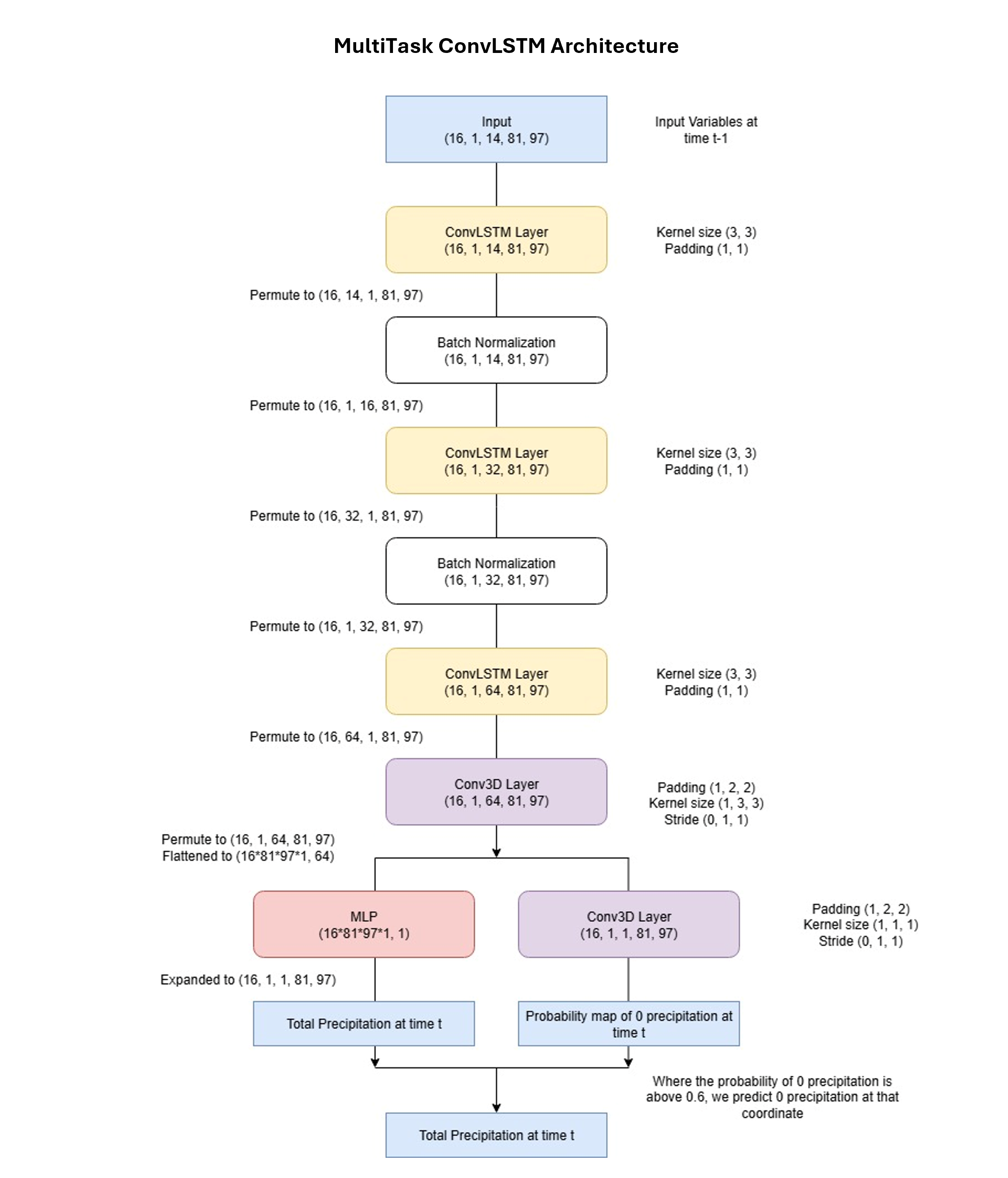}
    \caption*{\textbf{Figure S1. MultiTask ConvLSTM architecture.}  
    Schematic of the proposed architecture, with shared ConvLSTM layers and dual heads for classification and regression. The model takes as input hourly variables over latitude $-15^\circ$ to $5^\circ$ and longitude $-75^\circ$ to $-50^\circ$. These inputs are processed through a shared three-layer ConvLSTM encoder with batch normalization. A Conv3D layer refines the spatiotemporal features before branching into two task-specific heads: (i) a multilayer perceptron (MLP) regression head that estimates total precipitation at time t, and (ii) a Conv3D classification head that outputs the probability of zero rainfall at each grid cell. The dual-task design allows accurate discrimination between dry and rainy conditions.}
    \label{fig:ConvLSTM_architecture}
\end{figure}

\begin{figure}[t]
    \centering

    \includegraphics[width=1\textwidth]{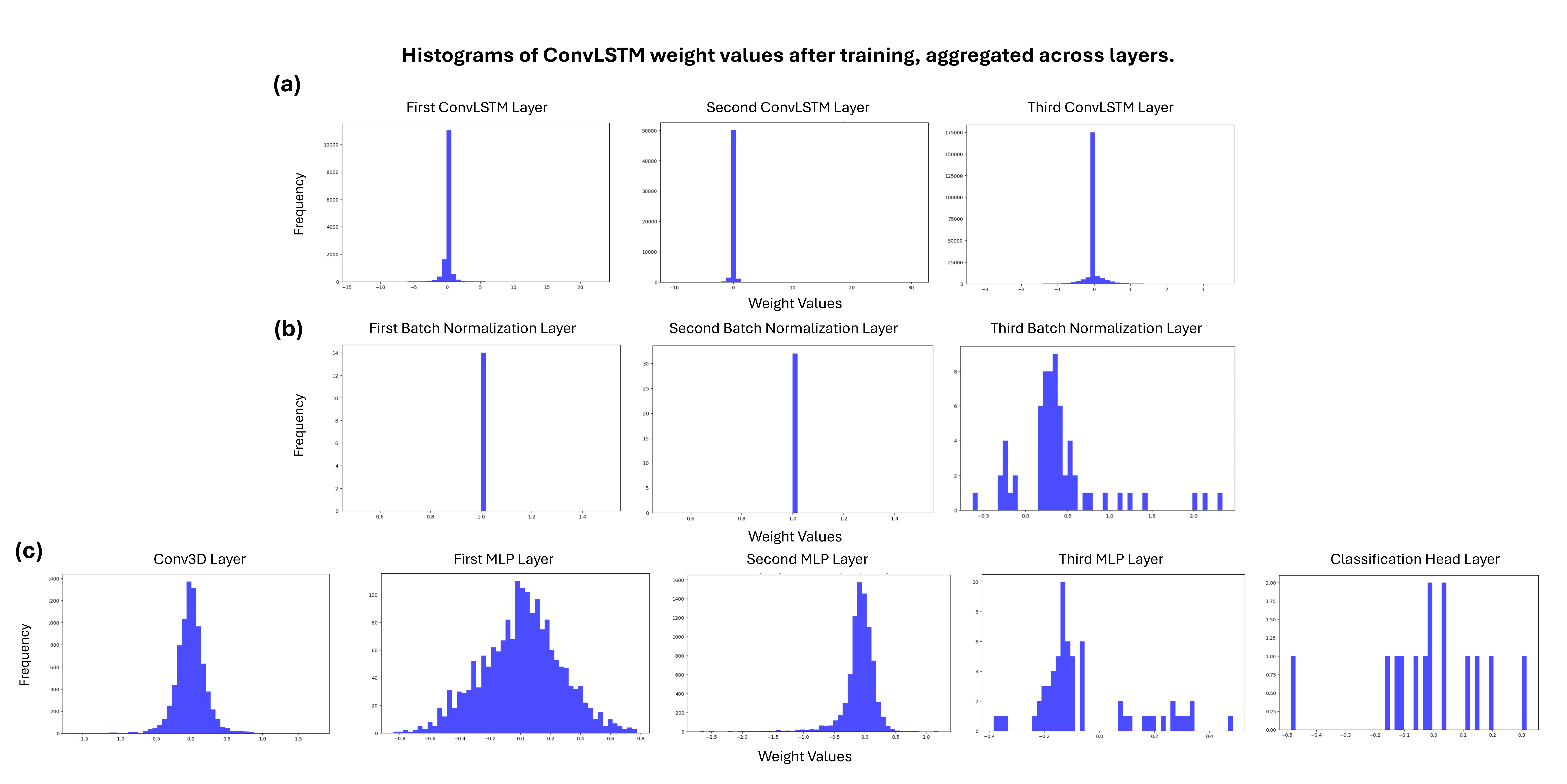}
    \caption*{\textbf{Figure S2. Histograms of learned weight distributions across layers in the MultiTask ConvLSTM, evaluated after training.} \textbf{(a)} ConvLSTM encoder layers. Weight distributions are sharply centered near zero, with broader spread across the first and third layers. \textbf{(b)} Batch Normalization layers. The first two layers display nearly degenerate distributions at unity, whereas the third exhibits greater variability, including stronger rescaling and shifting during training. \textbf{(c)} Output layers. The MLP layers show approximately Gaussian distributions, consistent with well-regularized training. The Conv3D layer and classification head display sparser distributions, with weights spread across a broader range.}
    \label{fig:ConvLSTM_weights}
\end{figure}

\begin{figure}[t]
    \centering

    \includegraphics[width=1\textwidth]{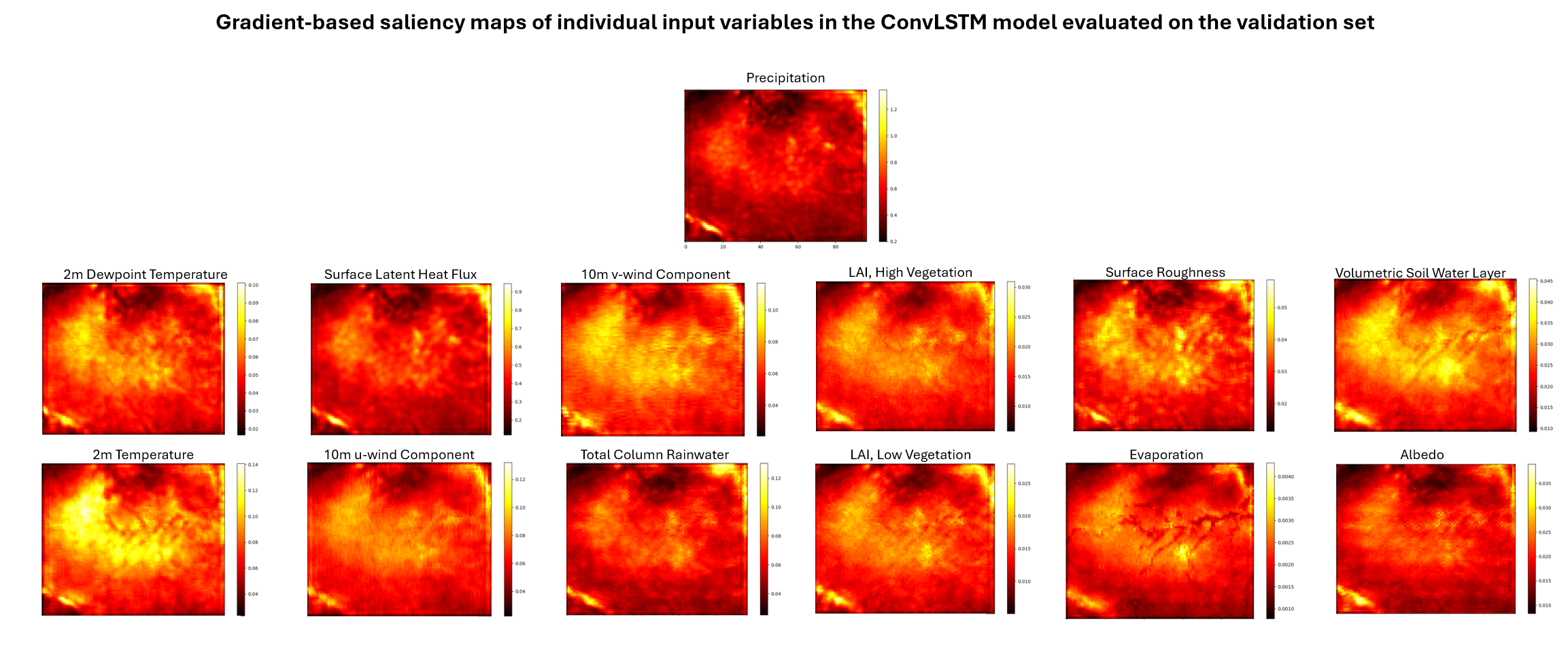}
    \caption*{\textbf{Figure S3. Gradient-based saliency maps of input variables in the MultiTask ConvLSTM, evaluated on the validation set.}  Each panel shows the spatial distribution of gradient magnitudes, with brighter intensities (yellow/white) indicating regions more influential for rainfall predictions. The model assigns the strongest sensitivity to the previous hour’s total precipitation, reflecting its role as the primary predictor of the spatial distribution and magnitude of subsequent rainfall. Vegetation variables display lower overall saliency, about an order of magnitude smaller than atmospheric inputs, consistent with their indirect role in short-term rainfall generation. Evaporation shows near-zero saliency, likely due to its strongly diurnal variability being captured via latent heat flux. Notably, vegetation saliency maps resemble the spatial structure of Amazonian biomass, with higher influence detected in ecologically sensitive regions such as the Andes, central Amazonia, and the Atlantic coastal margin. This suggests that vegetation inputs contribute spatially structured signals, particularly at biome boundaries where vegetation–atmosphere coupling is strongest.}
    \label{fig:ConvLSTM_saliencies}
\end{figure}

\FloatBarrier

\end{document}